\RequirePackage{rotating}
\documentclass[twocolumn]{aastex62}
\usepackage[utf8]{inputenc}
\usepackage[normalem]{ulem}
\usepackage{amsfonts,amsmath,amssymb}

\usepackage{xspace}
\usepackage{natbib}
\usepackage{verbatim}
\usepackage{graphicx,epstopdf}
\usepackage{fancyref}
\usepackage{hyperref}
\usepackage{breakurl}
\usepackage{amsmath}
\usepackage{graphicx} 
\usepackage{epstopdf}
\usepackage{graphicx}
\usepackage{appendix}
\usepackage{rotfloat}

\usepackage{amsmath}
\usepackage{graphicx}
\usepackage{natbib}
\usepackage{color}
\usepackage{rotating}
\usepackage{float}
\usepackage{booktabs}
\usepackage{graphicx}

\begin{document}
\title{A Systematic Analysis of Stellar Populations in the  Host Galaxies of Changing-look AGNs}

\author{Jun-Jie Jin}
\affiliation{Department of Astronomy, School of Physics, Peking University, Beijing 100871, People's Republic of China; E-mail: jjjin@pku.edu.cn; wuxb@pku.edu.cn}
\affiliation{Kavli Institute for Astronomy and Astrophysics, Peking University, Beijing 100871, People's Republic of China}
\author{Xue-Bing Wu}
\affiliation{Department of Astronomy, School of Physics, Peking University, Beijing 100871, People's Republic of China; E-mail: jjjin@pku.edu.cn; wuxb@pku.edu.cn}
\affiliation{Kavli Institute for Astronomy and Astrophysics, Peking University, Beijing 100871, People's Republic of China}
\author{Xiao-Tong Feng}
\affiliation{Department of Astronomy, School of Physics, Peking University, Beijing 100871, People's Republic of China; E-mail: jjjin@pku.edu.cn; wuxb@pku.edu.cn}
\affiliation{Kavli Institute for Astronomy and Astrophysics, Peking University, Beijing 100871, People's Republic of China}



\begin{abstract}
``Changing-look" active galactic nuclei (CL-AGNs) are a newly discovered class of AGNs that show the appearance (or disappearance) of broad emission lines within a short time scales (months to years) and are often associated with the dramatic changes of their continuum emissions.  They provide us with an unprecedented chance to directly investigate the host galaxy properties with minimal contamination from the luminous central engine during the ``turn-off" state, which is difficult for the normal luminous AGNs. In this work, for the first time, we systematically characterize the stellar populations and star formation histories (SFHs) of host galaxies for 26 turn-off CL-AGNs using the stellar population synthesis code STARLIGHT.  We find that the stellar populations of CL-AGNs are similar to those of normal AGNs, except that the intermediate stellar populations contribute more fractions. We estimate their stellar velocity dispersions ($\rm \sigma_{\star}$) and black hole masses ($\rm M_{BH,vir}$) and find that CL-AGNs also follow the overall $\rm M_{BH}-\sigma_{\star}$ relationship.  We also confirm the previous claims that CL-AGNs tend to be biased toward lower Eddington ratios, and that their extreme variabilities are more likely due to the intrinsic changes of accretion rates. In addition,  CL-AGNs with recent star formations (SF) tend to have higher Eddington ratios. Compared with previous studies, our analysis suggests that there may be a correlation between the CL-AGN host galaxy properties and their CL phenomena.

\end{abstract}


\section{Introduction} \label{sec:intro}

Active galactic nuclei (AGNs) are empirically classified into Type 1 and Type 2 according to their emission line features. Type 1 AGNs show both broad (FWHM $\sim$ 1000-10000 km/s) and narrow (FWHM $\lesssim$ 500km/s) Balmer emission lines, while only narrow emission lines are seen in Type 2 AGNs \citep{2015ARA&A..53..365N}. The broad emission lines are associated with the virialized, high-velocity gas in the broad line region (BLR) \citep{1988ApJ...325..114G,1989ApJ...345..637K,1991ApJS...75..719K,2003ApJ...589L..21M} and the narrow emission lines are emitted from the gas located in the narrow line region (NLR) at larger distance from the central engine. The unification model \citep{1985ApJ...297..621A,1993ARA&A..31..473A} explains the differences in Type 1 and Type 2 AGNs by the observer's line-of-sight towards the AGNs.  In this scheme,  two types of AGNs are intrinsically the same.  An equatorial dust and gas torus-like structure obscures the BLR along the line of sight, while the NLR at further distance remains unobsured. So Type 1 AGNs are viewed face-on while Type 2 AGNs are viewed edge-on.   Despite the success of the unification model, there are arguments that at least some Type 2 AGNs are intrinsically lacking broad lines because of their inadequate accretion rates \citep{2010ApJ...714..115S,2012MNRAS.426.3225B,2014A&A...568A.108P}.

The recent discoveries of CL-AGNs may also indicate a more complex picture of AGN unification.  CL-AGNs are a newly-discovered  class of AGNs that transit between Type 1 and Type 2 or vice versa, featuring emerging or disappearing broad Balmer emission lines within time scales of months to years. These sources were first discovered serendipitously \citep{2015ApJ...800..144L}, and then with the systematic searches, involving the repeated spectroscopic observations \citep{2016ApJ...821...33R,2019ApJ...874....8M,2018ApJ...862..109Y,2019ApJ...887...15W,2019ApJ...883L..44G}, large optical photometric variability \citep{2016MNRAS.457..389M,2019ApJ...883...31F} and mid-infrared (MIR) variability \citep{2018ApJ...864...27S,2018ApJ...866...26A,2020ApJ...889...46S}, the number of CL-AGNs has increased rapidly in recent years. Up to now, there are about $\rm \sim 100$ CL-AGNs have been discovered. Some of them have also been observed to change type more than once. For example, Mrk 1018 changed from Type 1.9 to Type 1 and later returned to Type 1.9 \citep{1986ApJ...311..135C,2016A&A...593L...8M}. Mrk 590 has changed from Type 1.5 to Type 1 and then to Type 1.9, as seen from monitoring over a period of more than 40 yr \citep{2014ApJ...796..134D}.  NGC 4151 was originally a Type 1.5 AGN \citep{1984MNRAS.211P..33P,1977ApJ...215..733O}, but its broad emission lines once disappeared and then appeared again \citep{2010A&A...509A.106S}. 

The physics origin of CL-AGNs is still not fully understood.  The main plausible mechanisms include : 1) the variable accretion rate, where an AGN follows an evolutionary sequence from Type 1 to an intermediate type and later to Type 2 (or vice versa), as the accretion rate decreasing \citep{2014MNRAS.438.3340E}; 2) variable obscuration, where the BLR is obscured due to the movement of the obscuring materials \citep{1989ApJ...342..224G,2012ApJ...747L..33E}; and 3)  tidal disruption events (TDEs), where a star is disrupted by a supermassive black hole (SMBH), leading to a bright flare \citep{2015MNRAS.452...69M}.  But with the increasing number of identified CL-AGNs, the rapid change in the black hole (BH) accretion rate is more preferred than other two  mechanisms. Much observational evidence has disfavored the variations in obscuration, such as the short transition timescale \citep{2015ApJ...800..144L}, the lag between the changes in the continuum and broad line emission \citep{2019ApJ...883...94T}, the large variability in mid-infrared \citep{2017ApJ...846L...7S} and low  polarization during ``turn-off" states \citep{2019A&A...625A..54H}. There are also many observation results for CL-AGNs that are not expected for a single TDE. For example, the lack of characteristic emission line features \citep{2016MNRAS.455.1691R,2018arXiv181103694K,2019ApJ...874....8M,2019ApJ...883...31F}, light-curve shapes different from those of TDEs \citep{2018arXiv181103694K}, the long-lived flares \citep{2016MNRAS.455.1691R,2019ApJ...874....8M} and the alternation between turn-on and off states \citep{2010A&A...509A.106S,2014ApJ...796..134D}. Though the dramatic changes of accretion rate are a more likely explanation, the standard viscous accretion disk model has failed to predict the CL-AGN transition timescale, and several alternative models have been proposed \citep{2018MNRAS.480.3898N,2019ApJ...883...76R,2019MNRAS.483L..17D,2019arXiv190406767S}. 

The CL phenomenon of CL-AGNs not only is important for understanding the structure and physics of accretion disks, but it can also provide an ideal case for investigating their host galaxies. For normal AGNs, the luminosity from AGNs is overwhelming compared with their host galaxies. Both the AGN continuum and broad emission lines seriously weaken the stellar features. However, for CL-AGNs, the faded central engines during the ``turn-off" states provide us an excellent chance to investigate the host galaxies in detail, with minimal contamination from the luminous nuclei. Moreover, the connection between small-scale BHs and large-scale host galaxies has been proposed for a long time \citep{2000ApJ...539L..13G,2001AIPC..586..363K,2001MNRAS.320L..30M,2002ApJ...574..740T}. The SMBHs commonly exist in galaxy centers, and the masses of these present-day SMBHs  are correlated with the properties of the host galaxies (such as bulge masses and velocity dispersions) \citep{1998AJ....115.2285M,2000ApJ...539L...9F}.  Many works also show that the evolutions of SMBHs (traced by AGNs) and the stellar populations of galaxies (traced by star-formation) are strikingly similar \citep{2004ApJ...615..209H,2008ApJ...679..118S,2010MNRAS.401.2531A}. 

Recently, some groups have already studied the host galaxies of CL-AGNs. \cite{2019ApJ...876...75C} characterized the host galaxies of four faded CL-AGNs using the broadband optical imaging, and suggested that the hosts of  CL-AGNs are predominantly disturbed or merging galaxies that lie in the ``blue cloud" along with the non-changing-look AGN (NCL-AGN) hosts in the color-magnitude diagram. For the first time, \cite{2019MNRAS.486..123R} studied the central region of the host galaxy of CL-AGN Mrk 590 by using high spatial resolution optical and near-infrared integral field unit (IFU) spectroscopy. They found that the average stellar population age is $>$ 5 Gyr in the central 10 arcsec, while $<$ 5 Gyr at larger radii (r $>$ 5 arcsec). Mrk 590 also has an $\rm {H\alpha}$ ring which shows a younger stellar population with an average age of 1-2 Gyr, and the gas dynamics suggest a complex balance between the gas inflow and outflow in the central region.  \cite{2020MNRAS.498.3985Y} identified five CL-AGNs in the MaNGA survey and cross-matched them with the GALEX-SDSS-WISE Legacy Catalog (GSWLC) to obtain their global star-formation rates (SFRs) and stellar masses.  They found that most CL-AGNs do possess pseudo-bulge features  and reside in the star-formation main sequence (SFMS), similar to the MaNGA NCL-AGNs. In a recent work, \cite{2021ApJ...907L..21D} showed that the CL-AGN hosts are mainly located in the green valley. All these results indicate that there may exist a potential link between the host environments of CL-AGNs and their changing-look phenomena. While these studies suggest that the host galaxy properties can be the key to understanding the environments that influence the BH growth and provide more insights into CL phenomena of CL-AGNs, there has been no large systematic statistical study of CL-AGN hosts. Previous results were mainly based on the image data, and the detailed information about CL-AGN hosts is still unclear. The only research with IFU data for a single source (Mrk 590) also lacks statistical significance, as one cannot draw strong conclusions from the study of a single object. In addition, the correlation between CL-AGN hosts and their central AGNs has not been studied yet.

In this work, we systematically analyze the ``turn-off" spectra of CL-AGN host galaxies in detail for the first time. With the help of stellar population synthesis code STARLIGHT, we determine the stellar populations and (star formation histories) SFHs of the host galaxies, and via the subtraction of stellar component, estimate the basic parameters of central BHs. The goal of this work is to obtain the detailed stellar populations of CL-AGN hosts and find clues on the connection between the host environments and CL phenomena.

This paper is organized as follows. Section 2 describes the sample selection and spectroscopic data. Section 3 outlines the method for spectral analysis. The output results of properties of CL-AGN hosts and their central BHs are given in Section 4. We discuss the connection between the CL-AGN host galaxies and CL phenomena in Section 5 and give a summary of our results in Section 6. We adopt the cosmology parameter $\rm H_{0}=70$ km $\rm s^{-1}$ $\rm Mpc^{-1}$ and a flat universe with $\rm \Omega_{M}=0.3$ and $\rm \Omega_{\Lambda}=0.7$.

\section{Data and Sample}

\subsection{Sample selection: ``turn-off" CL-AGNs}

As mentioned before, the ``turn-off" CL-AGNs are excellent cases for investigating the host galaxies in detail. In this section we mainly describe how to build a ``turn-off" spectral sample of CL-AGNs. 

We collect as many CL-AGNs that have been reported in the literature as possible. These sources are then screened according to following criteria:

1) Objects are required to have ``turn-off" spectra from Sloan Digital Sky Survey (SDSS) \footnote{The only exception is for object taken on MJD 58080 (J0159+0033), whose ``turn-off" spectrum was observed by eBOSS with the 2$\rm ^{\prime\prime}$ diameter fiber.}.  Because the spectra obtained from different telescopes  may have flux calibration offsets, which brings uncertainties to the analysis results, we require that all ``turn-off" spectra come from the same telescope, and SDSS is indeed a first choice. SDSS uses a 2.5 m wild-field telescope \citep{2006AJ....131.2332G} at Apache Point Observatory, and the observations were carried out by the fiber-field double spectrographs \citep{2013AJ....146...32S} with 3$\rm ^{\prime\prime}$ diameter fibers that resulted in acquiring more emissions from host galaxies. The reduced one-dimensional spectra have a wavelength coverage of 3800-9200 $\rm \AA$ at a spectral resolution $\rm R \sim 1500-2500$.  The wavelength coverage and spectral resolution of SDSS spectra meet the requirements for host galaxy stellar population analysis. Given all these reasons, we select the CL-AGNs with SDSS ``turn-off" spectra as our sample.

2) Objects must be Seyfert galaxies or quasars that changed to Type $\ge$ 1.9 during their ``turn-off" states. Intermediate types (such as 1.2 or 1.5) during the off states are abandoned because there may be still a large contribution from central AGNs, which will affect the spectral analysis of host galaxy. The Balmer emission line series in the optical regime are generally used to classify Seyfert types. Type 1 AGNs show broad components in all Balmer series, while Type 2 AGNs show only narrow components. The intermediate types show varying levels of broad component emissions \citep{1976MNRAS.176P..61O,1977ApJ...215..733O,1981ApJ...249..462O}. According to this definition, we require that the objects in our sample show no visible ${\rm H\beta}$ broad component at ``turn-off" states.

3) The redshift is no more than 0.35.  There are two reasons for this restriction: (1) to reject the high-redshift objects, as their AGNs are more luminous which may dilute the stellar features of host galaxies; (2) this redshift range not only covers the absorption lines needed for spectral analysis, but also avoids the ${\rm MgII}$ emission lines that may affect the spectral analysis when they enter the optical band. 

There are in total 26 objects in our sample. Table~\ref{table:observation} shows the observational properties of these objects.

\begin{table*}[]
	\caption{Observational information of 26 CL-AGNs.}
	\tiny
	\begin{tabular}{lccccccccccccc}
		\hline 
		\hline 
		name       &{Reference}\footnote{ References :  (1):\cite{2016ApJ...826..188R}; (2):\cite{2017ApJ...843..106B}; (3):\cite{2016MNRAS.457..389M}; (4):\cite{2019ApJ...883...31F}; (5):\cite{2018ApJ...862..109Y}; (6):\cite{2017ApJ...835..144G}; (7):\cite{2007arXiv...1811..03694}.}  & R.A.      & Decl.       & z &PLATE-MJD-FIBERID \footnote{The PLATE+MJD+FIBERID of ``turn-off" spectra, which uniquely determines a single observation of an object.}& \multicolumn{2}{c}{Obs. Date}    & \multicolumn{2}{c}{Instrument\footnote{The spectra of SDSS, eBOSS  and LAMOST are observed by fibers; The spectra of DCT, XLT, P60, Keck and LDSS3 are observed by slits.}}    & \multicolumn{2}{c}{Aperture}    \\	
		& & (J2000)  & (J2000)   &   &   & off  & on  & off   & on   & off   &on \\
		\\
		(1)        & (2)     & (3)     & (4)   & (5) &(6)  & \multicolumn{2}{c}{(7)}    & \multicolumn{2}{c}{(8)}    & \multicolumn{2}{c}{(9)}    \\
		\hline 
		J0126-0839   &(1)   &01:26:48   &-08:39:48   & 0.198   &2878-54465-0377   & 2008-02-02   & 2007-12-31   & SDSS   & SDSS   & $3.0^{\prime\prime}$   &$3.0^{\prime\prime}$ \\
		J0158-0052   &(2)   &01:58:05   &-00:52:22   & 0.080   &1075-52933-0140   & 2003-10-21   & -   & SDSS   & -   & $3.0^{\prime\prime}$   &- \\
		J0159+0033   &(3)   &01:59:58   &+00:33:11   & 0.312   &9384-58080-0605   & 2017-11-23   & 2000-11-23   & eBOSS   & SDSS   & $2.0^{\prime\prime}$   &$3.0^{\prime\prime}$ \\
		ZTF18aaabltn   &(4)   &08:17:26   &+10:12:10   & 0.046   &2423-54149-0301   & 2007-02-18   & 2019-05-02   & SDSS   & DCT   & $3.0^{\prime\prime}$   &$1.5^{\prime\prime}$ \\
		J0831+3646   &(5)   &08:31:32   &+36:46:17   & 0.195   &0827-52312-0064   & 2002-02-07   & 2015-12-11   & SDSS   & LAMOST   & $3.0^{\prime\prime}$   &$3.3^{\prime\prime}$ \\
		J0909+4747   &(5)   &09:09:32   &+47:47:31   & 0.117   &0899-52620-0340   & 2002-12-12   & 2016-12-23   & SDSS   & LAMOST   & $3.0^{\prime\prime}$   &$3.3^{\prime\prime}$ \\
		J0937+2602   &(5)   &09:37:30   &+26:02:32   & 0.162   &2294-53733-0616   & 2005-12-29   & 2015-12-13   & SDSS   & LMAOST   & $3.0^{\prime\prime}$   &$3.3^{\prime\prime}$ \\
		J1003+3525   &(5)   &10:03:23   &+35:25:04   & 0.119   &1951-53389-0400   & 2005-01-19   & 2017-04-24   & SDSS   & XLT   & $3.0^{\prime\prime}$   & $1.8^{\prime\prime}$ \\
		J1104+6343   &(5)   &11:04:23   &+63:43:05   & 0.164   &2882-54498-0588   & 2008-02-02   & 2002-04-06   & SDSS   & SDSS   & $3.0^{\prime\prime}$   &$3.0^{\prime\prime}$ \\
		J1110-0003   &(5)   &11:10:25   &-00:03:34   & 0.219   &0279-51984-0319   & 2001-03-16   & 2017-04-21   & SDSS   & XLT   & $3.0^{\prime\prime}$   & $1.8^{\prime\prime}$ \\
		J1115+0544   &(5)   &11:15:37   &+05:44:50   & 0.090   &0835-52326-0454   & 2002-02-21   & 2006-01-06   & SDSS   & LAMOST   & $3.0^{\prime\prime}$   &$3.3^{\prime\prime}$ \\
		J1132+0357   &(5)   &11:32:29   &+03:57:29   & 0.091   &0837-52642-0133   & 2003-01-03   & 2006-01-05   & SDSS   & LAMOST   & $3.0^{\prime\prime}$   &$3.3^{\prime\prime}$ \\
		ZTF18aasuray   &(4)   &11:33:56   &+67:01:07   & 0.040   &0492-51955-0273   & 2001-02-15   & 2018-06-21   & SDSS   & DCT   & $3.0^{\prime\prime}$   &$1.5^{\prime\prime}$ \\
		ZTF18aasszwr   &(4)   &12:25:50   &+51:08:46   & 0.168   &0971-52644-0417   & 2003-01-05   & 2018-12-03   & SDSS   & P60   & $3.0^{\prime\prime}$   &-\footnote{The final spectrum was defined by the brightest 70\% of the spaxels, so it's hard to verify the exact value of the aperture size \citep{2018PASP..130c5003B}. } \\
		ZTF18aahiqf   &(4)   &12:54:03   &+49:14:52   & 0.067   &1279-52736-0019   & 2003-04-07   & 2018-04-11   & SDSS   & DCT   & $3.0^{\prime\prime}$   &$1.5^{\prime\prime}$ \\
		J1259+5515   &(5)   &12:59:17   &+55:15:07   & 0.199   &1039-52707-0372   & 2003-03-09   & 2017-04-20   & SDSS   & XLT   & $3.0^{\prime\prime}$   & $2.3^{\prime\prime}$ \\
		J1319+6753   &(5)   &13:19:31   &+67:53:55   & 0.166   &0496-51973-0419   & 2001-03-05   & 2017-04-24   & SDSS   & XLT   & $3.0^{\prime\prime}$   & $2.3^{\prime\prime}$ \\
		J1358+4934   &(5)   &13:58:56   &+49:34:14   & 0.116   &1670-53438-0061   & 2005-03-09   & 2008-03-28   & SDSS   & SDSS   & $3.0^{\prime\prime}$   &$3.0^{\prime\prime}$ \\
		J1447+2833   &(5)   &14:47:54   &+28:33:24   & 0.163   &2141-53764-0134   & 2006-01-29   & 2015-02-18   & SDSS   & LAMOST   & $3.0^{\prime\prime}$   &$3.3^{\prime\prime}$ \\
		ZTF18aajupnt   &(4)   &15:33:08   &+44:32:08   & 0.037   &1052-52466-0450   & 2002-07-11   & 2018-08-07   & SDSS   & Keck   & $3.0^{\prime\prime}$   &$1.0^{\prime\prime}$ \\
		J1533+0110   &(5)   &15:33:56   &+01:10:30   & 0.143   &0363-51989-0575   & 2001-03-21   & 2008-04-05   & SDSS   & SDSS   & $3.0^{\prime\prime}$   &$3.0^{\prime\prime}$ \\
		J1545+2511   &(5)   &15:45:30   &+25:11:28   & 0.117   &1849-53846-0056   & 2004-06-16   & 2017-05-18   & SDSS   & LAMOST   & $3.0^{\prime\prime}$   &$3.3^{\prime\prime}$ \\
		J1550+4139   &(5)   &15:50:17   &+41:39:02   & 0.220   &1053-52468-0117   & 2002-07-13   & 2017-04-21   & SDSS   & XLT   & $3.0^{\prime\prime}$   & $1.8^{\prime\prime}$ \\
		J1552+2737   &(5)   &15:52:58   &+27:37:28   & 0.086   &1654-53498-0603   & 2005-05-08   & 2014-03-06   & SDSS   & LAMOST   & $3.0^{\prime\prime}$   &$3.3^{\prime\prime}$ \\
		J1554+3629   &(6)   &15:54:40   &+36:29:52   & 0.237   &1681-53172-0127   & 2004-06-16   & 2017-04-19   & SDSS   & XLT   & $3.0^{\prime\prime}$   & $2.3^{\prime\prime}$ \\
		PS1-13cbe   &(7)   &22:21:54   &+00:30:54   & 0.124   &1035-52816-0430   & 2003-06-26   & 2013-10-25   & SDSS   & LDSS3   & $3.0^{\prime\prime}$   &$1.0^{\prime\prime}$ \\
		\hline 
	\end{tabular}
	\label{table:observation}
\end{table*}

\subsection{Corresponding ``turn-on" Spectroscopy}  

As we mentioned before, we take the SDSS spectra to represent the ``off-state" spectra of CL-AGNs.  In order to  estimate the parameters of the central BHs during their ``turn-on" states, the ``turn-on" spectra are also needed. 

The ``turn-on" spectra are mainly based on the spectra from SDSS and LAMOST. LAMOST is a quasi-meridian reflecting Schmidt telescope equipped with an effective light-collecting aperture that varies from 3.6 to 4.9 m. It has 4000 fibers with a 5$\rm ^{\circ}$ field of view \citep{2012RAA....12.1197C,2012RAA....12..723Z}. The fiber diameter is $\rm 3^{\prime\prime}$ and the wavelength coverage ranges from 3700 to 9000 $\rm \AA$ with two arms \citep{2016ApJS..227...27D}, a blue arm (3700-5900 $\rm \AA$) and a red arm (5700-9000 $\rm \AA$). The spectral resolution of LAMOST is approximately R $\rm \sim 1800$ over the entire wavelength range. The data we used are reduced by LAMOST pipelines \citep{2012RAA....12.1243L} and included in LAMOST quasar catalog \citep{2016AJ....151...24A,2018AJ....155..189D,2019ApJS..240....6Y}.

In addition, we also collect some ``turn-on" spectra from published papers \citep{2019ApJ...883...31F,2007arXiv...1811..03694}.   The ``turn-on" spectra of  ZTF18aaabltn, ZTF18aasuray and ZTF18aahiqf were observed by the Deveny spectrograph on the  Discovery Channel Telescope with a  $\rm 1.5^{\prime\prime}$ wide slit and a wavelength coverage of 3600-8000 $\rm \AA$. The ZTF18aajupn was observed by Keck I LRIS \citep{2019ApJ...883...31F}. The long-slit spectroscopic observations of J1003+3525, J1259+5515, J1319+6753, J1550+4139 and J1554+3629 \citep{2018ApJ...862..109Y} were carried out using the Xinglong 2.16 m telescope (XLT) equipped with the Beijing faint Object Spectrograph and Camera (BFOSC), and yield a resolution of R $\rm \sim $ 265 or 340\footnote{The observations were carried out with a $\rm 1.8^{\prime\prime}$ slit or a $\rm 2.3^{\prime\prime}$ slit when the  seeing is less or larger than $\rm < 2^{\prime\prime}$.} at 5000 $\rm \AA$ with a wavelength coverage ranges from 3850 to 8300 $\rm \AA$ \citep{2016PASP..128k5005F}. The ``turn-on" spectrum of PS1-13cbe was observed by using the Low Dispersion Survey Spectrograpg-3 (LDSS3) on the 6.5 m Magellan Clay telescope \citep{2018arXiv181103694K}. The $\rm 1^{\prime\prime}$-wide long slit with the VPH-all grism yields a wavelength coverage of 3700-10000 $\rm \AA$ with a resolution of $\rm \sim$ 9 $\rm \AA$. The observation of ZTF18aasszwr ``turn-on" spectrum was carried out using the Spectral Energy Distribution Machine (SEDM; \citealt{2018PASP..130c5003B}) IFU spectrograph on the Palomar 60-inch (P60; \citealt{2006PASP..118.1396C}) operating as part of ZTF. At last, we collected the ``turn-on" spectra for 25 sources in our sample (only J0158-0052 doesn't have the corresponding ``turn-on" spectrum).  The observational information for all objects during both the ``turn-on" and ``turn-off" states are summarized in Table~\ref{table:observation}.

The spectra from LAMOST and other telescopes need to be calibrated to SDSS spectra to obtain a homogeneous set of spectra with the same flux calibration. Assuming narrow emission lines remain constant during the monitoring period, a particularly robust technique to correct the flux calibration offset is to match the strength of narrow emission lines, such as $ \rm {[OIII] \lambda 5007}$. The basic idea for this assumption is that the forbidden lines arise in an extended ($\rm \sim 10^{2-3} $pc) region, so the light-travel times and recombination times are very long (hundreds of years).  Studies on CL-AGNs also support this assumption. \cite{2017ApJ...835..144G} found that the $\rm {[OIII]\lambda5007}$ luminosity during ``turn-on" states is consistent with that measured during ``turn-off" states.  \cite{2018ApJ...858...49W} compared the  $\rm {[OIII]\lambda5007}$ emission line profiles in CL-AGN SDSS J141324.27+530527.0 taken at four different epochs, and their results showed an extremely high consistence. 

However, there are two potential difficulties in using this method. (1):The narrow emission lines adopted for scaling AGN spectra might be difficult to isolate in some cases, and might be blended with time-varying features. An effective way to deal with this problem is to check the difference obtained after subtracting one spectrum from another. If the spectra are not correctly scaled, residual features will appear at the position of any constant line. It is tricky to adjust the relative scaling factor until the residual narrow-line features in the difference spectrum vanish \citep{2006pces.conf...89P}. \cite{1992A&A...266...72W} found that the relative flux calibration accuracy can be achieved at the $\rm \sim 2 \%$ level.  (2):  The observed NLR surface brightness distribution may be larger than the aperture size, which can affect the narrow-line-based flux calibration. However, it is hard to correct the deviation of scaling factors in our work because both the intrinsic NLR surface brightness distribution for our sample and the seeing conditions for some observations are unknown.  We estimated the projected angular sizes of the NLR based on the radius-luminosity relation \citep{2013ApJ...774..145H,2013MNRAS.430.2327L}, and compared them with the  size of teh spectroscopic apertures. The comparison showed that only 3 objects need additional aperture corrections because their NLR sizes are larger than the aperture sizes used during the on-state observations. Even the estimated upper-limit values\footnote{Assuming the [OIII]$\lambda$5007 has a uniform distribution, the upper-limit of the scaling factor of cross-calibration is calculated by multiplying the factor $\rm NLR\_size/A_{on}$, where the $\rm NLR\_size$ is the estimated projected angular size of NLR based on the radius luminosity relation \citep{2013MNRAS.430.2327L,2013ApJ...774..145H}, and $\rm A_{on}$ is the aperture size used during on-state observations.} of the scaling factor did not affect our results and conclusions, so we didn't apply any additional aperture corrections to the flux calibration in our work. At last, the ``turn-on" spectra were thus scaled to the SDSS spectra. This allows us to obtain a relatively homogeneous set of spectra with the same flux calibration. The scaling factors of five sources with SDSS ``turn-on" spectra all float around 1.00 (0.98, 1.00, 1.10, 0.70, and 1.10, respectively), thus validating our method being reliable.

\begin{figure*}[!htb]
	\begin{center}
		\includegraphics[angle=0,scale=0.4,keepaspectratio=flase]{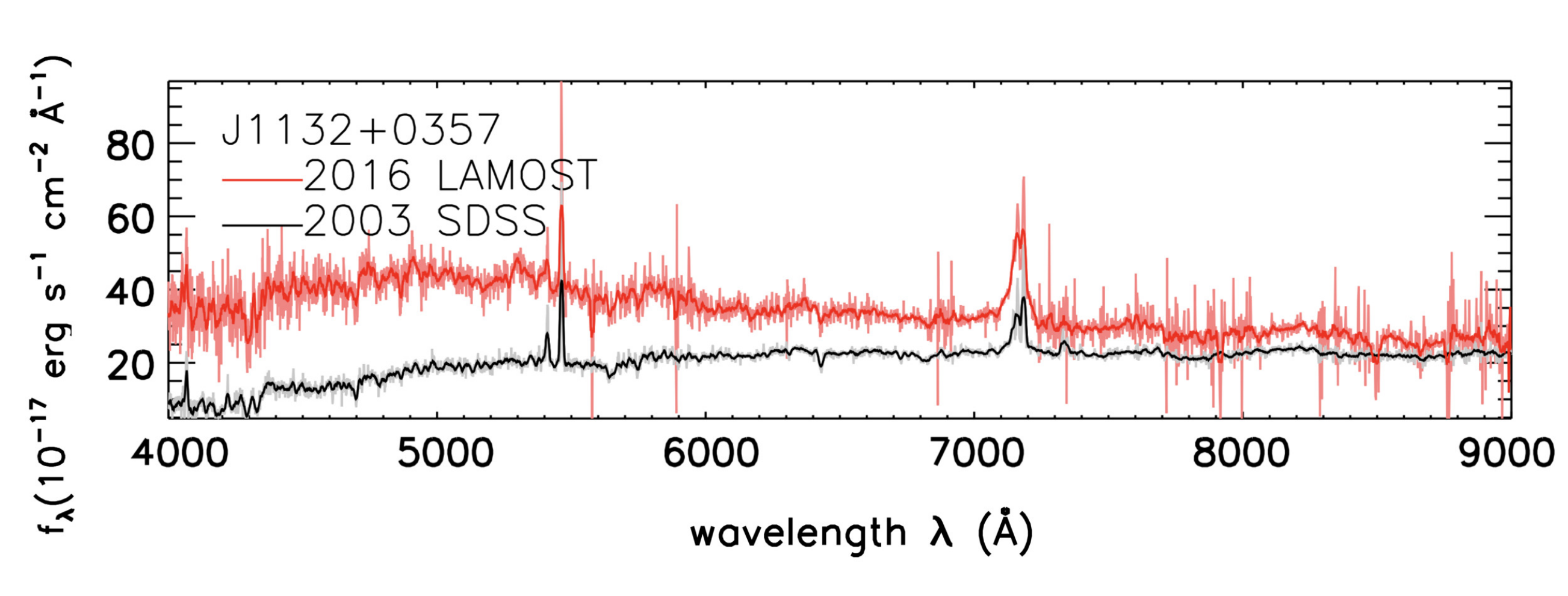}
		\caption{An example of  comparison between teh ``turn-on" (red) and ``turn-off" (black) spectra. The object comes from \cite{2018ApJ...862..109Y}. No corrections for the redshift or Galactic extinction. The spectra of total 26 CL-AGNs are shown in Appendix.\label{fig:Spec}}
	\end{center}
\end{figure*}

In total, 23 ``turn-on" spectra were flux calibrated. The spectral resolution of  ZTF18aasszwr is not high enough to distinguish the signatures of narrow emission lines. In another case, ZTF18aasuray, where the $ \rm {[OIII] \lambda 5007}$ scaling yields an unphysical result by displaying a significant flux decrease in the long wavelength range after the flux cross-calibration. In these two cases, we instead use the originally observed spectra.  Figure ~\ref{fig:Spec} shows an example of the ``turn-on" and ``turn-off" spectra of CL-AGN J1132+0357. The spectral comparisons for allot the CL-AGNs in the sample are shown in the Appendix Figure ~\ref{fig:Specs}.

\section{Spectral Fitting}

In this section, we present the stellar population synthesis analysis of the CL-AGN host galaxies, and describe the decomposition of the spectra into host galaxy and AGN components.  Prior to fittings, all spectra were deredshifted to the rest frame and corrected for the Galactic extinction using the maps from \cite{1998ApJ...500..525S} and the extinction curves from \cite{1999PASP..111...63F}.

\subsection{Fitting for Host Galaxy Components}


\subsubsection{Spectral Synthesis with STARLIGHT}
We derive the stellar populations of CL-AGN hosts from ``turn-off" spectra using the spectral analysis code  STARLIGHT \citep{2005MNRAS.358..363C}. This code uses a mixture of simulated annealing plus Metropolis scheme and Markov Chain Monte Carlo techniques to yield the minimum $\rm \chi^{2}$ value ($\rm \chi^{2} = \sum_{\lambda}[(O_{\lambda}-M_{\lambda})\omega_{\lambda}]^2$), where  $\rm O_{\lambda}$ is the observed spectrum and the $\rm {\omega_{\lambda}}^{-1}$ is the error in $\rm O_{\lambda}$ at each wavelength bin. The model spectrum $\rm M_{\lambda}$ ($\rm \vec{x}$, $\rm A_{V}$, $\rm v_{\star}$, $\rm \sigma_{\star}$) has $\rm N_{\star}+3$ parameters: the linear combination of $\rm N_{\star}$ Simple Stellar Populations (SSPs) from the evolutionary synthesis models, teh extinction $\rm A_{V}$, the velocity shift $\rm v_{\star}$, and velocity dispersion $\rm \sigma_{\star}$. The $\rm \chi^{2}/N_{\lambda}$ that we use in Section ~\ref{sec:STARLIGHT} is the fitted $\rm \chi^{2}$ divided by the number of wavelength bins used in the fit. 

\subsubsection{Parameter Determination and Stellar Population} \label{sec:STARLIGHT}

During the fitting, we adopt the spectral bases with 21 ages (ranging from 1 Myr to 10 Gyr ) to provide enough resolution in age. The base spectra are normalized at $\rm \lambda_{0} = 4020 \AA$, and the observed spectra are normalized to the median flux between 4010 $\rm \AA$ and 4050 $\rm \AA$. Although some of the CL-AGN ``turn-off" spectra do not show an obvious quasar-like featureless continuum (FC), it is still possible that their spectra contain an  AGN component.  So beside the stellar populations, a power-law component for an AGN continuum ($\rm F_{\lambda} \varpropto \lambda^{\alpha_{\lambda}}$ ) is added in the fittings if needed.  

\begin{figure}[!htb]
	\begin{center}
		\includegraphics[angle=0,scale=0.36,keepaspectratio=flase]{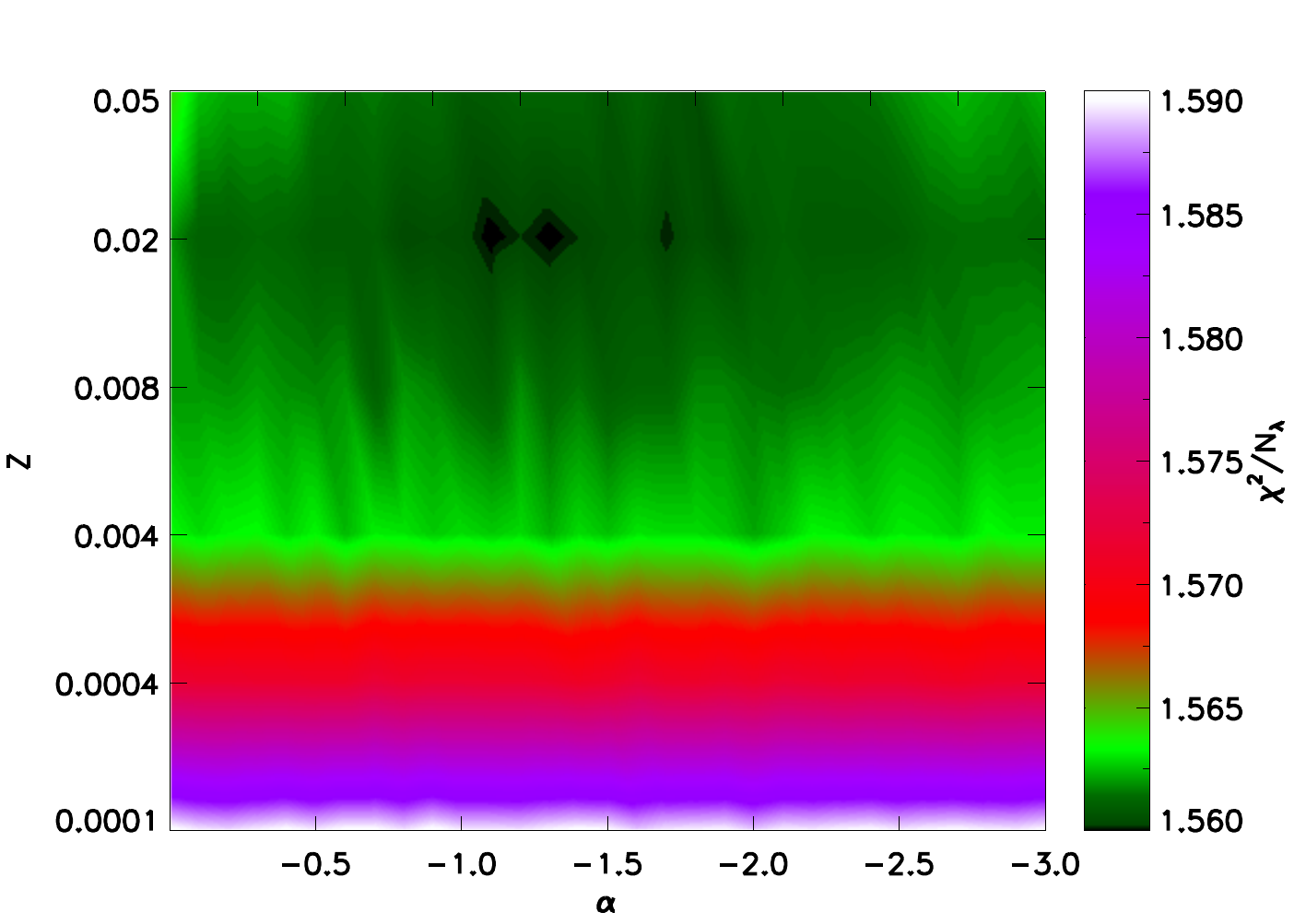}
		\caption{ An example of the averaged $\rm \chi^{2}/N_{\lambda}$ over 100 individual fittings with different seeds and artificial spectra of object J1104+6343. Each fitting adopts a certain metallicity (Z) and a power-law index ($\rm \alpha$), as well as 25 SSPs. \label{fig:AZ}}
	\end{center}
\end{figure}

In order to minimize the parameter space and make the results more reliable, we calculate the best-fit metallicity ($\rm Z$) and power-law index ($\rm \alpha_{\lambda}$) by using the method proposed by \cite{2010ApJ...718..928M}.  We vary the value of  $\rm \alpha_{\lambda}$ from -3.0 to 0.0 at intervals of 0.5 to search for the best fit power-law index (non-power-law fittings represented by $\rm \alpha_{\lambda}=0$).  To search for the best-fit metallicity, we carry out a metallicity test with 6 metallicities (Z = 0.0001, 0.0004, 0.004, 0.008, 0.02, and 0.05). During the fitting, the weights ($\rm \omega_{\lambda}$) of strongest stellar absorption features ($\rm Ca \,II \,H\&K$ and G band) are increased by multiplying 2, and the emission lines are masked and excluded from the fitting.  In order to evaluate the uncertainties in fitting, we perform a series of Monte Carlo simulations. In each fitting, we generate an artificial spectrum by adding Gaussian random errors with the matching signal-to-noise ratio (S/N).  We then fit these artificial spectra with the fitting procedure described above.  We evaluate the fitting quality by the minimum  $\rm \chi^{2}/N_{\lambda}$ value, and average it over 100 times by fitting with different seeds (the random numbers that need to be appointed during the STARLIGHT fitting) and artificial spectra. Figure ~\ref{fig:AZ} shows an example for the fitting results with different power-law indices and metallicities. The $\rm \chi^{2}/N_{\lambda}$  value reaches its minimum at $\rm Z=0.02$ and $\rm \alpha=-1.3$, and we take them as the best-fit metallicity and power-law index.

\begin{figure*}[!htb]
	\begin{center}
		\includegraphics[angle=0,scale=1.,keepaspectratio=flase]{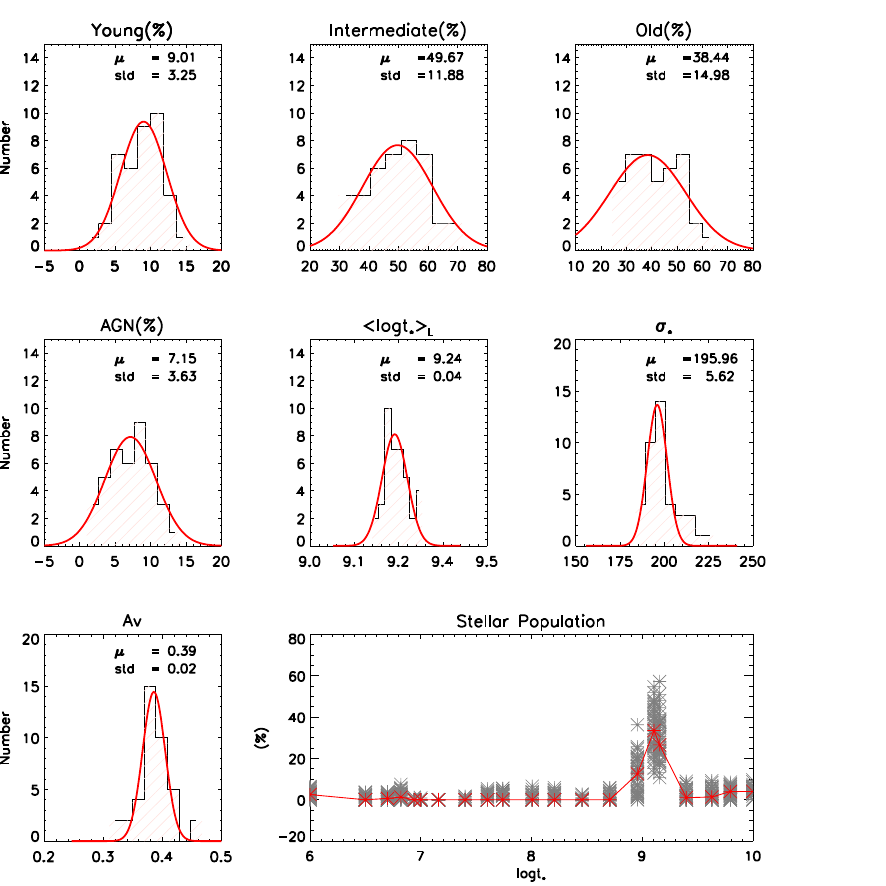}
		\caption{ An example of obtaining uncertainties of STARLIGHT fitting parameters for object J0831+3646. We perform 100 fits with different seeds and artificial spectra. The mean values of the resulting parameters and their standard deviations from Gaussian fits (red solid lines) are given on the top of each panel.  The first seven panels show the percentages of young, intermediate and old stellar components, AGN component, mean age weighted by light, stellar velocity dispersion and intrinsic extinction, respectively.  The last panel shows the light contribution from different stellar populations. The black crosses represent the fitting over 100 times and the red points give the average values of each stellar population. \label{fig:Noise}}
	\end{center}
\end{figure*}

\begin{figure*}[!htb]
	\begin{center}
		\includegraphics[angle=0,scale=0.7,keepaspectratio=flase]{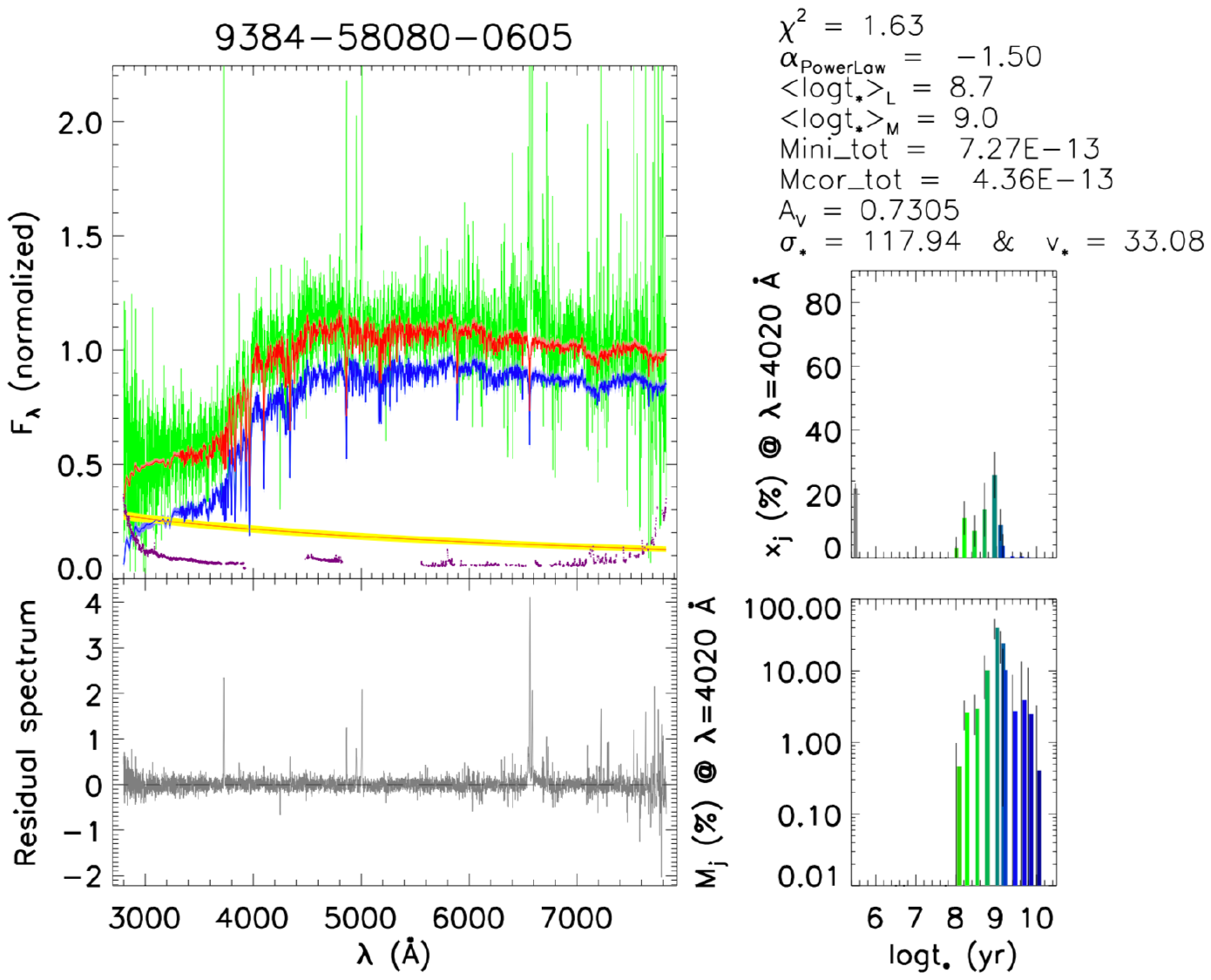}
		\caption{An example of the spectral synthesis of a CL-AGN J0159+0033 during the ``turn-off" state.  Top left: the observed spectrum $\rm O_{\lambda}$ (green) and its error (purple), the best-fit model spectrum $\rm M_{\lambda}$ (red), the stellar component (blue) and the AGN power law (yellow). The solid lines represents the mean results and the shaded regions represent the standard deviations; Bottom left: the residual spectrum $\rm E_{\lambda}$ (gray); The light-weighted stellar population fraction $\rm x_{j}$ is shown in the top right panel,  where the gray bar on the left represents the contribution from the AGN. The black line on each bar represents the standard deviation of each component in the fitting. Similarly, the mass-weighted population fraction $\rm \mu_{j}$ is shown in the bottom right panel. Other parameters, such as the power-law index, the mean age (weighted by light $\rm <logt_{\star}>_{L}$ and by mass $\rm <logt_{\star}>_{M}$), $\rm \sigma_{\star}$, and $\rm A_{V}$, are also shown in the top right corner. \label{fig:Spec_1}}
	\end{center}
\end{figure*}

There are in total 100 fitting results for the given metallicity and power-law index.  Though the differences occur, the uncertainties due to the random seed and spectral noise do not significantly change the overall population distributions. Figure ~\ref{fig:Noise} shows that the outcomes for each stellar contribution  and different parameters vary slightly ($\rm \lesssim 10\%$). We take the mean value for each parameter, and estimate the uncertainty by the standard deviation. The final results of the STARLIGHT fitting for J0159+0033 are summarized in Figure ~\ref{fig:Spec_1}, as an example.

\subsubsection{SFH: The UV-to-optical Luminosity in the Past}

SFH is one of the principal markers of galaxy growth. It can reveal the evolutions of galaxies in quantitative detail and provide an evolutionary pictures that can be compared with other evolutionary sequences, such as the evolutions of BHs. Since the stellar populations in CL-AGN hosts are already resolved, we can calculate the luminosity of the stellar components from UV to optical over the past $\rm \sim$1 Gyr by reconstructing the UV-to-optical spectrum. The calculations are described in detail in  \cite{2010ApJ...718..928M}, and are briefly mentioned here.

Since the ``turn-off'' spectra were observed by the $\rm 3^{\prime\prime}$ diameter fiber, aperture correction must be performed to obtain the precise luminosity from the whole galaxy. The aperture correction for the stellar component is derived from equation (9) in  \cite{2010ApJ...718..928M}. It should be noted that this method assumes that the emissions from stellar populations detected through the fiber are uniformly distributed over the galaxy. Issues such as the galaxy orientation and the patchy distributions of stellar populations can bring uncertainties in the aperture correction estimation \citep{2003ApJ...599..971H}.


The rest-frame spectrum $\rm F_{i}(\lambda,t)$ from UV to optical (912  $\rm  \sim $ 9000 $\rm \AA$) can be recovered after the correction for the intrinsic extinction by using equation (6) in  \cite{2010ApJ...718..928M}: 

\begin{equation}
\rm F_{i}(\lambda,t)=Mcor\_tot \sum_{j=1}^{N}\frac{\mu_{j}}{f_{\star,j,t_{0}}}f_{\star,j,t}B_{\lambda,j,t} , 
\end{equation}

\begin{figure}[!htb]
	\begin{center}
		\includegraphics[angle=0,scale=0.3,keepaspectratio=flase]{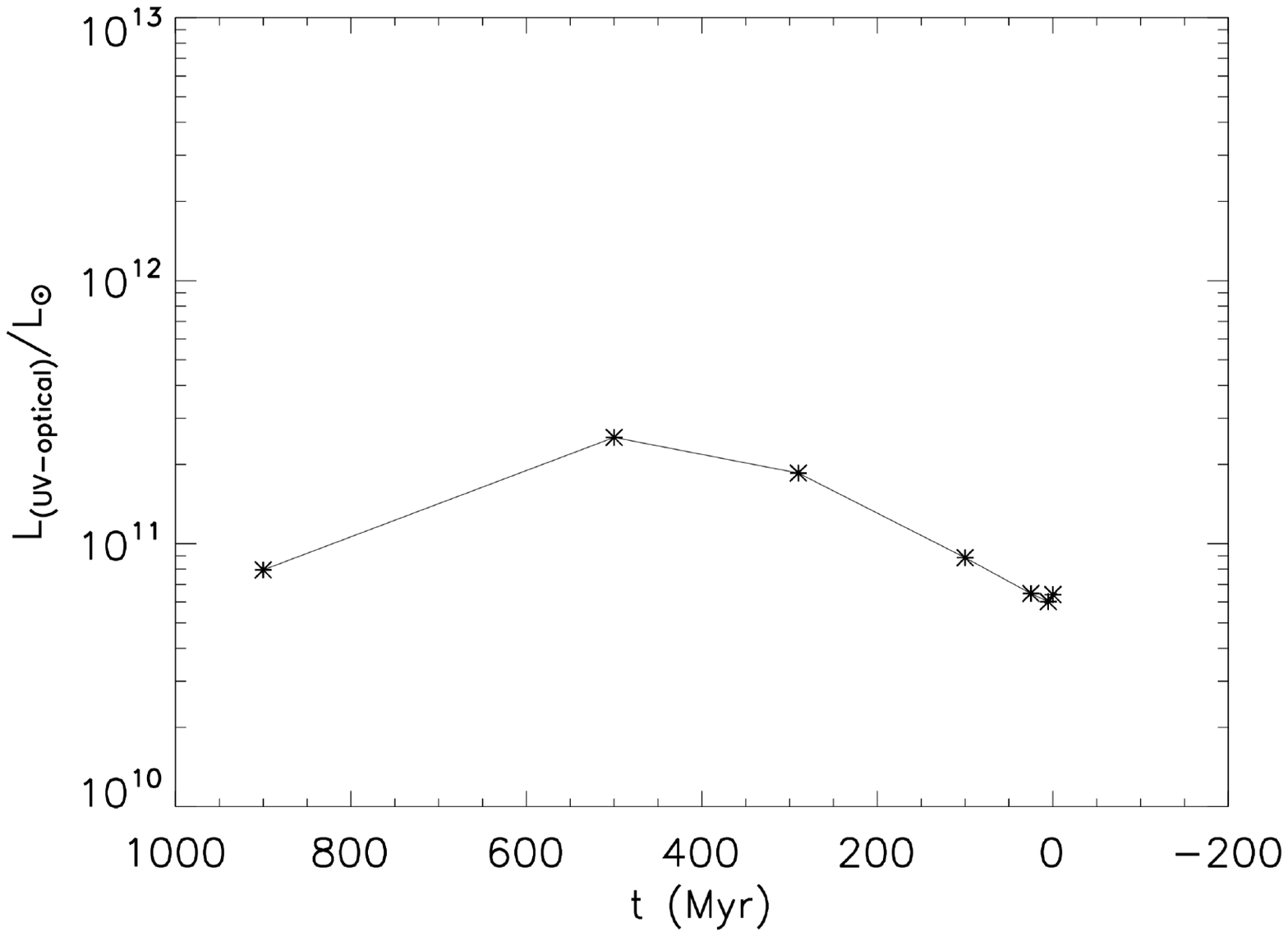}
		\caption{ An example of SFH of object J0126-0839, which shows $\rm L_{(UV-optical)}$ in the past 0, 25, 100, 290, 500 and 900 Myr respectively. \label{fig:SFH}}
	\end{center}
\end{figure}

~\\
\noindent where $\rm F_{i}(\lambda,t)$ is the rebuilt stellar spectrum at time $\rm t$. $\rm Mcor\_tot$ is the present stellar mass after aperture correction, $\rm \mu_{j}$ is the mass-weighted fraction for different stellar components, $\rm B_{\lambda,j,t}$ is the base of the SSPs from \cite{2003MNRAS.344.1000B}, abbreviated as BC03 SSP, without normalization, $\rm f_{\star,j,t_{0}}$ is the present ($\rm t_{0}$) fraction of the remaining stellar mass to the initial mass of population $\rm j$, and $\rm f_{\star,j,t}$ is such a fraction at time $\rm t$. Detailed descriptions about these parameters can be found in \cite{2005MNRAS.358..363C}. After integrating \rm $F_{i}(\lambda,t)$  from 912 $\rm \AA$ to 9000 $\rm \AA$, we estimate the UV-to-Optical luminosity ($\rm L_{UV-optical}$) of the past 0, 25, 100, 290, 500 and 900 Myr separately. An example of past stellar luminosity during $\rm \sim$ 1 Gyr is shown in Figure ~\ref{fig:SFH}. In this case, there is an obvious starburst (SB) around $\rm 500 Myr$.

\subsubsection{AGN Featureless Continuum versus Young stellar Components}  \label{sec:degeneracy}

A common question often arises when fitting the stellar components of AGN host galaxies: the degeneracy between SBs and the AGN FC in galaxies that harbor active nuclei. It is hard to distinguish between the emissions from the AGN power-law and young SBs due to the narrow wavelength coverage of our data. The main difference between these two components lies in the Balmer absorption lines and the blue side of Balmer jump. But these wavelength ranges are masked due to the strong emission lines from the central AGN or are affected by the high noise due to the falling of detector efficiency in the blue end of the spectra. As a result, the blue continuum can be attributed to either young stellar populations,  the AGN FC or a combination of both.

\begin{figure}[!htb]
	\begin{center}
		\includegraphics[angle=0,scale=0.32,keepaspectratio=flase]{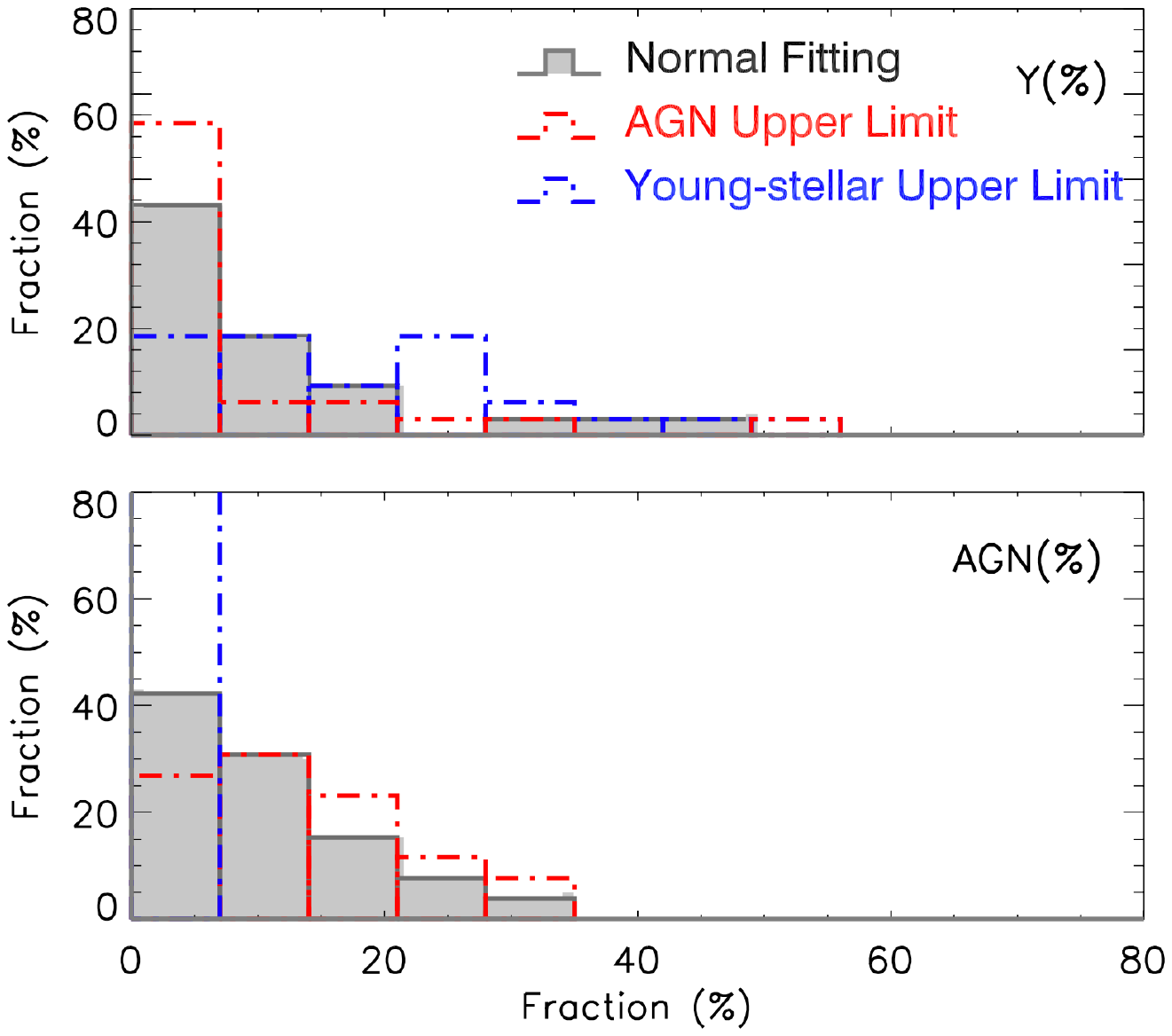}
		\caption{ The distributions of $\rm Y(\%)$ (top panel) and $\rm AGN(\%)$ (bottom panel) among 3 different cases. Y(\%) means the light fraction from the young stellar component, and AGN (\%) means the light fraction from AGN. The detailed descriptions can be found in Section~\ref{sec:degeneracy}. The gray histograms represent the distribution calculated from normal fitting; The red histograms represent the results fitted by the base spectra that do bot contain young stellar populations (regarded as the upper limit of AGN fraction); and the blue histograms represent the results fitted by the base spectra that do not contain the AGN power law (regarded as the upper limit of the young stellar fraction). The total numbers of each histogram add up to 100. \label{fig:Fraction}}
	\end{center}
\end{figure}

While there is no way of breaking this SB-FC degeneracy without the broader spectral coverage, the uncertainty of the STARLIGHT fitting results due to this degeneracy can be estimated. To do this, we adopt two extreme cases to fit the spectra: 1) estimate the upper-limit of AGN emission contribution. In this case the extremely young stellar populations ($\rm age \le 5Mr$) are not included during the fitting, and the fitting results of AGN fraction, AGN (\%), are regarded as the upper limit values of AGN contribution; 2)  estimate the upper-limit of young stellar population emission contribution. In this case the AGN power-law continuum is not included during the fitting, and such results of young stellar populations fraction, Y (\%), are regarded as the upper limit values of young stellar population contribution. Figure ~\ref{fig:Fraction} shows a comparison between these two extreme cases  and the normal fitting results as described in section 3.1.1 and 3.1.2. As we can see, the two extreme conditions do have an impact on the fitting results. When compared with the normal fitting results, both upper limits are slightly shifted toward the higher fraction. In the following analysis, we will show that these deviations do not affect the final conclusions. What's more, the objects in the spectral synthesis results without AGN components (ZTF18aasszwr, ZTF18aahiqf and J1554+3629) show absences of not only $\rm {H\beta}$ broad emission lines, but also $\rm {H\alpha}$ broad emission lines in the ``turn-off" spectra, thus validating the results of the STARLIGHT synthesis.

\subsection{Spectral properties of CL-AGNs}

It is essential to remove the stellar component to isolate the ``pure" AGN spectrum before analyzing the AGN continuum and emission lines, and the variations of AGN properties between two epochs can be used. In this section, we subtract the host galaxy emissions from the spectra of both epochs, then fit the AGN components and analyze the properties of each source in our sample. 

\subsubsection{AGN Spectral Fitting}

\begin{figure}[!htb]
	\begin{center}
		\includegraphics[angle=0,scale=0.3,keepaspectratio=flase]{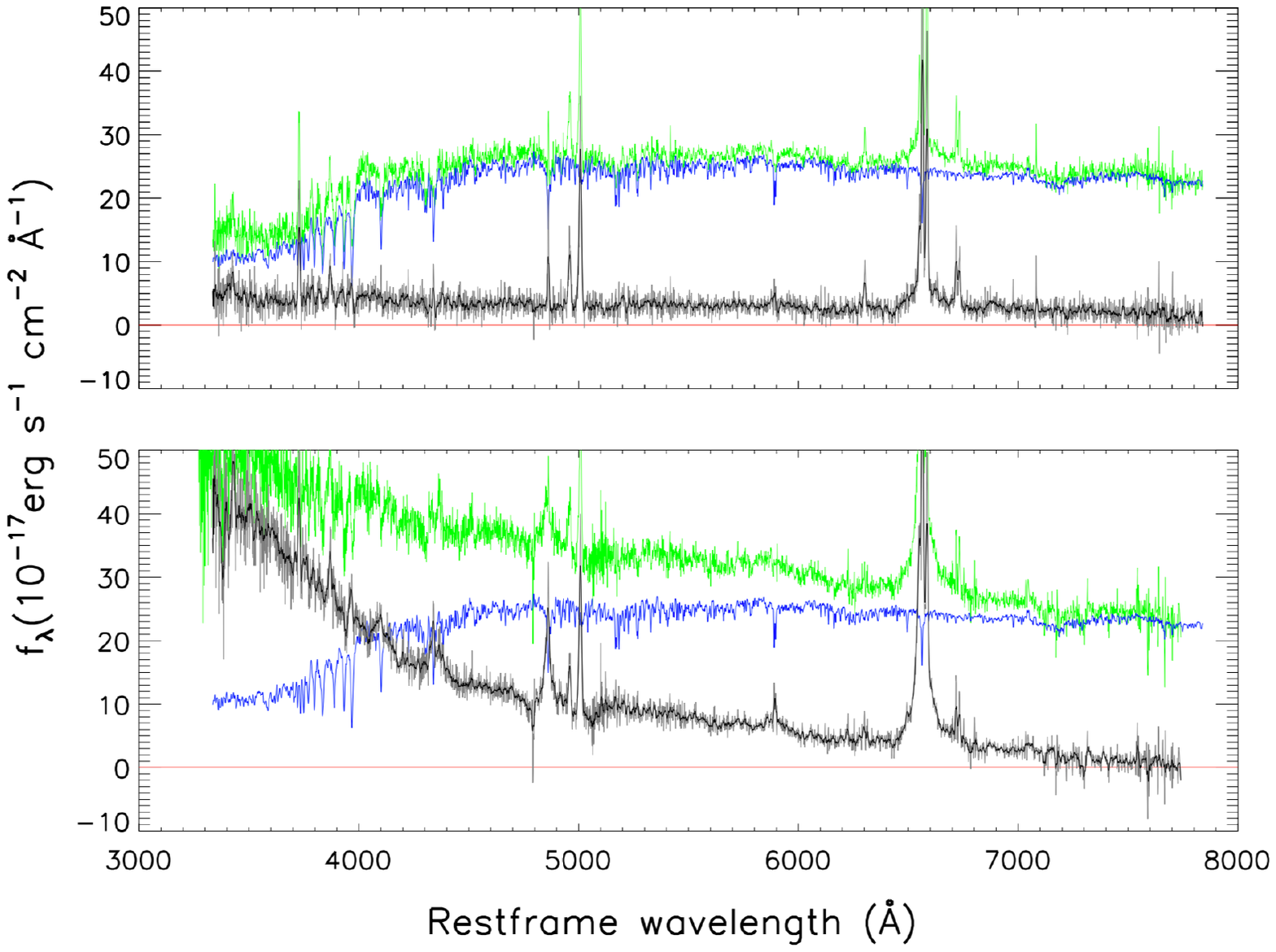}
		\caption{ An example of the decomposition of ``turn-off" (upper panel) and ``turn-on" (lower panel) spectra of object J1447+2833. The green line in each panel shows the observed spectra in the rest frame. The blue line represents the best-fit stellar component from the ``turn-off" spectra, and the black line represents the pure emission from AGN. \label{fig:HostSub}}
	\end{center}
\end{figure}

The contamination from host galaxies can't be negligible for quasars. Especially for the $\rm z \lesssim 0.5$ low-luminosity quasars \citep{2011ApJS..194...45S}, the host contamination is on average $\rm \sim 15\%$, which leads to a $\rm \sim 0.06$ dex overestimation of the $\rm 5100 \AA$ continuum luminosity ($\rm L_{5100}$). Unfortunately, the quality of  most of the ``turn-on" spectra does not allow for a reliable galaxy continuum subtraction, since the stellar features are overwhelmed by the extremely bright emissions from AGN. But for CL-AGNs, the stellar component can be well constrained by fitting the ``turn-off" spectra, which are dominated by the host galaxy continuum and the absorption lines, such as $\rm {Ca}$ $\rm {H}$ and $\rm {K}$ transitions. The host galaxy emission is not expected to change within the few decades between the observations. To isolate the pure AGN emissions from ``turn-on" spectra, we freeze the host component to the best-fit stellar model found from SDSS ``turn-off" spectra. Figure ~\ref{fig:HostSub} shows an example result of the decomposition of ``turn-off'' spectra and ``turn-on" spectra. 

\begin{figure*}[!htb]
	\begin{center}
		\includegraphics[angle=0,scale=0.73,keepaspectratio=flase]{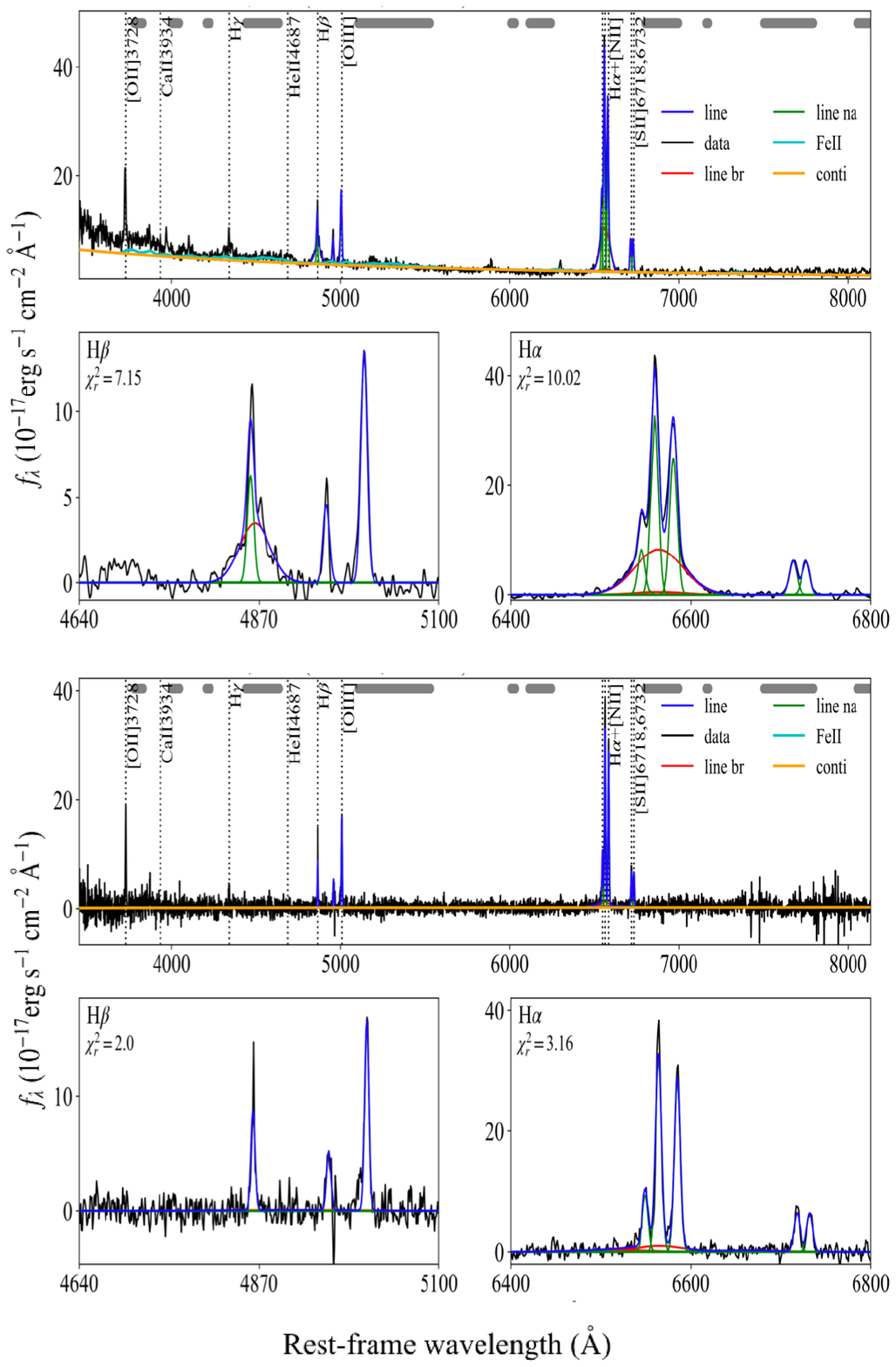}
		\caption{Examples for the multi-component spectral decompositions of the ``turn-on" (up) and ``turn-off" (below) spectra of object J2221+0030. For each spectrum, the upper subplot shows the original spectrum (black), Fe II emissions (cyan) and AGN power-law continuum (yellow), respectively. The best-fit AGN emission lines are in blue, which is the sum of narrow lines and broad lines. The bottom subplot shows the detailed emission components of $\rm {H\beta}$ and $\rm {H\alpha}$ lines. The broad components are in red while the narrow ones are in green, along with their sum (blue).\label{fig:Emission}}
	\end{center}
\end{figure*}

After subtracting the best-fit stellar component from the observed spectra of both epochs, we fit the spectra of each object using the publicly available multicomponent spectral fitting code PyQSOFit \citep{2018ascl.soft09008G}. A detailed description of the code and its application can be found in  \cite{2018ascl.soft09008G} and \cite{2019ApJS..241...34S}. Using the decomposed quasar spectra in both epochs for each object in our sample, we measure the properties of $\rm {H\alpha}$ and $\rm {H\beta}$ (both narrow and broad) emission lines as well as $\rm L_{5100}$. We fit the AGN continuum with a power law and Fe II model, and add a polynomial component if necessary. For the Balmer emission lines, we use multiple ($\rm 1\sim3$) Gaussians to account for the broad component ($\rm FWHM \ge 1200 km/s$) and one single Gaussian to account for the narrow component ($\rm FWHM < 1200 km/s$).  To evaluate the uncertainty during the fitting, we generate 50 mock spectra by adding Gaussian noise to the original spectrum. The measured properties of emission lines and $\rm L_{5100}$ are tabulated in Table~\ref{table:sample21} and Table~\ref{table:sample22}, and the examples of spectral fittings are presented in Figure ~\ref{fig:Emission}.

\begin{sidewaystable*}[htbp]
	~\\
	~\\
	~\\
	~\\
	~\\
	~\\
	~\\
	~\\
	~\\
	\caption{The parameters of central BHs in different states and the fitting results of broad emission lines.}
	\setlength{\tabcolsep}{3.5pt}
	\label{table:sample21}
	\tiny
	\centering
	\begin{tabular}{lccccccccccccc}
		\hline
		\hline
		 \multicolumn{1}{c}{name} & $\rm F_{H\alpha } (broad)$ &$\rm FWHM{H\alpha } (broad)$&  $\rm F_{H\beta} (broad)$  &  $\rm FWHM{H\beta} (broad)$ & \multicolumn{2}{c}{$\rm L_{5100}$} & \multicolumn{2}{c}{$\rm L_{bol}$} & \multicolumn{2}{c}{$\rm log(Edd)$}& $\rm M_{BH,vir}$ &$\rm \triangle_{t}$(yr) \\
		 & $\rm 10^{-17} erg \,s^{-1} \,cm^{-2}$ & $\rm km \,s^{-1}$ & $\rm 10^{-17} erg \,s^{-1} \,cm^{-2}$  & $\rm km \,s^{-1}$ &  \multicolumn{2}{c}{$\rm log(erg \,s^{-1})$}& \multicolumn{2}{c}{$\rm log(erg \,s^{-1})$} & \multicolumn{2}{c}{$\rm log(erg \,s^{-1})$}  & $\rm log(M/M_{\odot})$& yr \\
		&  & &  & &  on & off\footnote{Although the analysis of STARLIGHT shows that there are no AGN components in ZTF18aasszwr, ZTF18aahiqh and J1554+3629, QSOFit still give the $\rm L_{5100}$ values. We keep these values and mark them with *.}  &  on & off &  on & off & & \\
		\\
		 \multicolumn{1}{c}{(1)} & (2) &(3) & (4) &(5)&\multicolumn{2}{c}{(6)} & \multicolumn{2}{c}{(7)} &  \multicolumn{2}{c}{(8)} &(9)&(10)\\
		\hline
		J0126-0839&	  $ 1049.05\pm   29.97$ &	   $ 4539.12\pm  150.29$ &	   $  318.13\pm   18.02$ &	   $ 4147.57\pm  292.76$ &	   $43.57\pm 0.01$ &	   $42.77\pm 0.03$ &	   $44.52\pm 0.01$ &	   $43.72\pm 0.03$ &	   $-0.17\pm 0.34$ &	   $-0.97\pm 0.34 $	 &  $ 7.70^{+ 0.11}_{- 0.10} $ &	   6.0&\\
J0158-0052&	  -- &	   -- &	   -- &	   -- &	   -- &	   $42.31\pm 0.01$ &	   -- &	   $43.26\pm 0.01$ &	   -- &	   $ 5.16\pm 0.01 $	 &  -- &	   11.0&\\
J0159+0033&	  $  923.80\pm   45.65$ &	   $ 3618.27\pm  182.33$ &	   $  338.57\pm    8.84$ &	   $ 5464.20\pm  280.01$ &	   $43.74\pm 0.01$ &	   $43.22\pm 0.01$ &	   $44.70\pm 0.01$ &	   $44.18\pm 0.01$ &	   $-0.63\pm 0.41$ &	   $-1.15\pm 0.41 $	&   $ 7.71^{+ 0.12}_{- 0.10} $ &	   9.0&\\
ZTF18aaabltn&	  $ 4523.12\pm  155.85$ &	   $ 3772.20\pm  231.51$ &	   $ 2029.12\pm   57.49$ &	   $ 4486.96\pm  156.61$ &	   $42.17\pm 0.01$ &	   $41.29\pm 0.10$ &	   $43.13\pm 0.01$ &	   $42.25\pm 0.11$ &	   $-1.65\pm 0.21$ &	   $-2.53\pm 0.23 $	 &  $ 7.13^{+ 0.13}_{- 0.11} $ &	   12.0&\\
J0831+3646&	  $  922.95\pm  135.93$ &	   $ 7004.81\pm  805.54$ &	   $  175.68\pm   40.57$ &	   $ 2972.20\pm  190.92$ &	   $42.85\pm 0.02$ &	   $42.39\pm 0.05$ &	   $43.81\pm 0.02$ &	   $43.35\pm 0.05$ &	   $-2.39\pm 0.12$ &	   $-2.85\pm 0.13 $	 &  $ 8.05^{+ 0.18}_{- 0.17} $ &	   14.0&\\
J0909+4747&	  $ 1690.95\pm  117.50$ &	   $ 3576.43\pm  264.21$ &	   $  327.90\pm   58.23$ &	   $ 3957.68\pm 1090.01$ &	   $42.96\pm 0.01$ &	   $42.54\pm 0.01$ &	   $43.92\pm 0.01$ &	   $43.50\pm 0.01$ &	   $-1.19\pm 0.22$ &	   $-1.61\pm 0.22 $	 &  $ 7.32^{+ 0.13}_{- 0.12} $ &	   14.0&\\
J0937+2602&	  $ 1738.48\pm  325.41$ &	   $ 2961.09\pm  476.65$ &	   $  237.79\pm   60.91$ &	   $ 2828.79\pm  646.70$ &	   $42.89\pm 0.04$ &	   $42.77\pm 0.01$ &	   $43.84\pm 0.04$ &	   $43.72\pm 0.01$ &	   $-1.41\pm 0.18$ &	   $-1.54\pm 0.17 $	 &  $ 7.33^{+ 0.20}_{- 0.19} $ &	   8.0&\\
J1003+3525&	  $ 3142.97\pm  147.97$ &	   $ 7606.87\pm  660.98$ &	   $ 3114.64\pm  132.51$ &	   $12701.87\pm  610.76$ &	   $42.86\pm 0.03$ &	   $42.54\pm 0.02$ &	   $43.81\pm 0.03$ &	   $43.50\pm 0.02$ &	   $-1.88\pm 0.14$ &	   $-2.19\pm 0.14 $	 &  $ 8.15^{+ 0.16}_{- 0.15} $ &	   12.0&\\
J1104+6343&	  $  830.70\pm   84.74$ &	   $ 5543.50\pm  204.87$ &	   $   76.63\pm    9.56$ &	   $ 4504.30\pm  871.82$ &	   $42.24\pm 0.02$ &	   $42.12\pm 0.06$ &	   $43.19\pm 0.02$ &	   $43.07\pm 0.06$ &	   $-1.97\pm 0.50$ &	   $-2.09\pm 0.51 $	 &  $ 7.72^{+ 0.12}_{- 0.10} $ &	   6.0&\\
J1110-0003&	  $  983.19\pm   34.67$ &	   $ 6242.31\pm  285.65$ &	   $  543.05\pm   33.82$ &	   $ 7142.73\pm  565.77$ &	   $43.30\pm 0.03$ &	   $43.12\pm 0.01$ &	   $44.25\pm 0.03$ &	   $44.07\pm 0.01$ &	   $-1.24\pm 0.23$ &	   $-1.42\pm 0.22 $	 &  $ 8.02^{+ 0.12}_{- 0.11} $ &	   16.0&\\
J1115+0544&	  $ 2484.03\pm  106.34$ &	   $ 3477.19\pm   78.44$ &	   $ 1087.24\pm   54.53$ &	   $ 4044.40\pm  244.85$ &	   $42.97\pm 0.01$ &	   $41.75\pm 0.07$ &	   $43.93\pm 0.01$ &	   $42.70\pm 0.07$ &	   $-1.36\pm 0.15$ &	   $-2.58\pm 0.17 $	  & $ 7.25^{+ 0.10}_{- 0.08} $ &	   14.0&\\
J1132+0357&	  $ 1297.51\pm   90.82$ &	   $ 3733.58\pm  308.00$ &	   $ 1191.77\pm   79.55$ &	   $ 7048.59\pm   10.14$ &	   $43.17\pm 0.02$ &	   $41.91\pm 0.04$ &	   $44.12\pm 0.02$ &	   $42.86\pm 0.04$ &	   $-1.28\pm 0.14$ &	   $-2.54\pm 0.15 $	  & $ 7.17^{+ 0.14}_{- 0.13} $ &	   13.0&\\
ZTF18aasuray&	  $ 6357.29\pm  821.67$ &	   $ 4191.15\pm  447.46$ &	   $ 2672.91\pm  178.04$ &	   $ 5162.87\pm  435.79$ &	   $43.00\pm 0.01$ &	   $41.59\pm 0.02$ &	   $43.95\pm 0.01$ &	   $42.54\pm 0.02$ &	   $-1.52\pm 0.13$ &	   $-2.93\pm 0.13 $	  & $ 7.24^{+ 0.16}_{- 0.15} $ &	   17.0&\\
ZTF18aasszwr&	  $13967.28\pm   30.40$ &	   $ 7498.52\pm   35.20$ &	   $ 4084.67\pm   21.78$ &	   $ 7722.92\pm   50.45$ &	   $44.24\pm 0.00\footnote{The error is less than 0.005.}$ &	   $42.06\pm 0.06 ^*$ &	   $45.20\pm 0.00$ &	   $43.01\pm 0.06 ^*$ &	   $-1.22\pm 0.19$ &	   $-3.40\pm 0.20 ^*$	&   $ 8.68^{+ 0.10}_{- 0.09} $ &	   16.0&\\
ZTF18aahiqf&	  $ 8489.94\pm  847.24$ &	   $ 9074.59\pm  571.92$ &	   $ 1060.01\pm  117.10$ &	   $ 7662.35\pm  665.68$ &	   $42.82\pm 0.01$ &	   $41.75\pm 0.04 ^*$ &	   $43.77\pm 0.01$ &	   $42.70\pm 0.04 ^*$ &	   $-2.12\pm 0.10$ &	   $-3.19\pm 0.11 ^*$	  & $ 8.26^{+ 0.15}_{- 0.13} $ &	   15.0&\\
J1259+5515&	  $ 1604.73\pm  248.54$ &	   $ 2874.68\pm  115.28$ &	   $  455.88\pm   31.57$ &	   $ 2824.73\pm    8.15$ &	   $43.07\pm 0.04$ &	   $42.52\pm 0.06$ &	   $44.02\pm 0.04$ &	   $43.48\pm 0.06$ &	   $-2.38\pm 0.19$ &	   $-2.93\pm 0.20 $	  & $ 7.39^{+ 0.11}_{- 0.10} $ &	   14.0&\\
J1319+6753&	  $  606.64\pm  145.08$ &	   $ 3247.77\pm  376.32$ &	   $  853.39\pm  240.09$ &	   $12186.68\pm 3976.47$ &	   $43.12\pm 0.03$ &	   $42.86\pm 0.02$ &	   $44.07\pm 0.03$ &	   $43.81\pm 0.02$ &	   $-1.11\pm 0.24$ &	   $-1.37\pm 0.24 $	  & $ 7.17^{+ 0.18}_{- 0.17} $ &	   16.0&\\
J1358+4934&	  $ 1242.98\pm   75.23$ &	   $ 3241.02\pm  260.92$ &	   $  324.86\pm   21.57$ &	   $ 5592.89\pm  712.74$ &	   $43.00\pm 0.01$ &	   $42.40\pm 0.02$ &	   $43.95\pm 0.01$ &	   $43.35\pm 0.02$ &	   $-0.41\pm 0.56$ &	   $-1.01\pm 0.56 $	  & $ 7.15^{+ 0.13}_{- 0.12} $ &	   3.0&\\
J1447+2833&	  $ 2114.86\pm  138.79$ &	   $ 3254.51\pm  178.09$ &	   $  305.27\pm   22.86$ &	   $ 2822.07\pm   19.61$ &	   $43.51\pm 0.01$ &	   $43.07\pm 0.00$ &	   $44.46\pm 0.01$ &	   $44.03\pm 0.00$ &	   $-1.16\pm 0.14$ &	   $-1.59\pm 0.14 $	 &  $ 7.46^{+ 0.12}_{- 0.10} $ &	   9.0&\\
ZTF18aajupnt&	  $ 1115.12\pm  182.71$ &	   $ 2895.42\pm   41.74$ &	   $  843.45\pm   17.13$ &	   $ 2824.82\pm    0.21$ &	   $41.09\pm 0.13$ &	   $40.97\pm 0.14$ &	   $42.04\pm 0.13$ &	   $41.92\pm 0.15$ &	   $-3.10\pm 0.20$ &	   $-3.22\pm 0.21 $	 &  $ 6.45^{+ 0.11}_{- 0.10} $ &	   16.0&\\
J1533+0110&	  $ 1034.48\pm   28.61$ &	   $ 3891.65\pm  151.29$ &	   $  224.91\pm   15.63$ &	   $ 3913.54\pm  313.04$ &	   $42.83\pm 0.01$ &	   $42.53\pm 0.02$ &	   $43.78\pm 0.01$ &	   $43.49\pm 0.02$ &	   $-1.64\pm 0.16$ &	   $-1.93\pm 0.16 $	  & $ 7.38^{+ 0.11}_{- 0.10} $ &	   7.0&\\
J1545+2511&	  $ 1623.30\pm   98.45$ &	   $ 3707.73\pm  151.07$ &	   $  312.26\pm   21.80$ &	   $ 2853.70\pm  126.13$ &	   $43.00\pm 0.01$ &	   $42.74\pm 0.01$ &	   $43.95\pm 0.01$ &	   $43.70\pm 0.01$ &	   $-1.73\pm 0.13$ &	   $-1.99\pm 0.13 $	 &  $ 7.34^{+ 0.11}_{- 0.10} $ &	   11.0&\\
J1550+4139&	  $  755.72\pm   91.77$ &	   $ 2829.66\pm    5.18$ &	   $  607.33\pm   60.48$ &	   $ 6909.07\pm 1355.49$ &	   $43.23\pm 0.07$ &	   $42.58\pm 0.05$ &	   $44.18\pm 0.07$ &	   $43.53\pm 0.05$ &	   $-1.91\pm 0.17$ &	   $-2.56\pm 0.16 $	  & $ 7.25^{+ 0.10}_{- 0.08} $ &	   15.0&\\
J1552+2737&	  $ 1658.80\pm   21.39$ &	   $ 8529.44\pm  262.64$ &	   $  388.91\pm   51.67$ &	   $ 7025.70\pm 2126.10$ &	   $42.26\pm 0.03$ &	   $41.89\pm 0.05$ &	   $43.21\pm 0.03$ &	   $42.84\pm 0.05$ &	   $-2.37\pm 0.18$ &	   $-2.74\pm 0.19 $	 &  $ 7.94^{+ 0.12}_{- 0.11} $ &	   9.0&\\
J1554+3629&	  $ 2555.38\pm  141.26$ &	   $ 4488.48\pm  430.77$ &	   $  699.36\pm   89.89$ &	   $ 5506.71\pm 1801.97$ &	   $43.86\pm 0.02$ &	   $41.98\pm 0.07 ^*$ &	   $44.82\pm 0.02$ &	   $42.93\pm 0.08 ^*$ &	   $-1.22\pm 0.18$ &	   $-3.11\pm 0.19 ^*$	&   $ 8.00^{+ 0.15}_{- 0.14} $ &	   12.0&\\
PS1-13cbe&	  $  605.75\pm    6.19$ &	   $ 2824.58\pm   22.27$ &	   $  168.62\pm    3.25$ &	   $ 2825.82\pm    0.21$ &	   $42.83\pm 0.00$ &	   $41.59\pm 0.09$ &	   $43.78\pm 0.00$ &	   $42.54\pm 0.09$ &	   $-0.29\pm 0.36$ &	   $-1.53\pm 0.37 $	 &  $ 6.89^{+ 0.10}_{- 0.08} $ &	   10.0&\\
		\hline
	\end{tabular}
	\tablecomments{Columns 2-5: Flux and FWHM of broad components measured from ``turn-on" epoch of CL-AGNs. Columns 6-8: The estimated $\rm L_{5100}$, $\rm L_{bol}$, and $\rm log(Edd)$, respectively. Column 9: Virial black hole mass derived from the $\rm R_{BLR}-L_{5100}$ relation. Column 10: The lag between the ``turn-on" and ``turn-off" epochs in CL-AGNs, which is calculated from the optical light curve, or observational time difference between two epochs (if there is no light-curve available, or light-curve coverage is not enough).}
\end{sidewaystable*}

\begin{sidewaystable*}
	\caption{The results of narrow lines profile modeling. }
	\setlength{\tabcolsep}{3.5pt}
	\label{table:sample22}
	\tiny
	\begin{tabular}{lcccccccccccc}
		\hline
		\hline
		&\multicolumn{2}{c}{$\rm F_{H\alpha } (narrow)$} & \multicolumn{2}{c}{$\rm F_{H\beta} (narrow)$}& \multicolumn{2}{c}{$\rm F_{[OIII] \lambda 5007} $}   &  \multicolumn{2}{c}{$\rm F_{[NII] \lambda 6585} $}  & \multicolumn{2}{c}{$\rm F_{[SIII] \lambda 6718} $ } &  \multicolumn{2}{c}{$\rm F_{[SIII] \lambda 6732} $} \\
		& \multicolumn{2}{c}{$\rm 10^{-17} erg \,s^{-1} \,cm^{-2}$} &  \multicolumn{2}{c}{$\rm 10^{-17} erg \,^{-1} \,cm^{-2}$} &  \multicolumn{2}{c}{$\rm 10^{-17} erg \,s^{-1} \,cm^{-2}$} &  \multicolumn{2}{c}{$\rm 10^{-17} erg \,s^{-1} \,cm^{-2}$} &  \multicolumn{2}{c}{$\rm 10^{-17} erg \,s^{-1} \,cm^{-2}$} &  \multicolumn{2}{c}{$\rm 10^{-17} erg \,s^{-1} \,cm^{-2}$} \\	
		& on & off  &  on & off &  on&  off &  on & off & on & off& on & off \\ 
		\\
		 \multicolumn{1}{c}{(1)}   &  \multicolumn{2}{c}{(2)}  & \multicolumn{2}{c}{(3) } &  \multicolumn{2}{c}{(4)} & \multicolumn{2}{c}{(5)} & \multicolumn{2}{c}{(6)} & \multicolumn{2}{c}{(7)}\\
		\hline
		J0126-0839&	 $414.27\pm  7.36$ &	   $105.53\pm  4.50$ &	   $ 81.32\pm  4.13$ &	   $ 70.12\pm  1.38$ &	   $ 72.51\pm  4.07$ &	   $ 67.65\pm  1.72$ &	   $188.19\pm  6.13$ &	   $144.48\pm  1.64$ &	   $ 57.28\pm  2.59$ &	   $ 39.02\pm  0.86$ &	   $ 57.40\pm  2.60$ &	   $ 39.10\pm  0.87$ \\
J0158-0052&	 -- &	  $394.52\pm  4.42$ &	   -- &	   $123.50\pm  3.42$ &	   -- &	   $274.46\pm  4.82$ &	   -- &	   $107.80\pm  2.35$ &	   -- &	   $ 75.24\pm  1.86$ &	   -- &	   $ 75.40\pm  1.87$ \\
J0159+0033&	 $190.51\pm  6.01$ &	   $200.26\pm  3.75$ &	   $ 51.23\pm  3.23$ &	   $ 49.22\pm  1.52$ &	   $109.64\pm  3.17$ &	   $ 98.20\pm  1.85$ &	   $ 94.53\pm  6.24$ &	   $106.66\pm  3.67$ &	   $ 52.72\pm  4.08$ &	   $ 48.35\pm  1.67$ &	   $ 52.83\pm  4.09$ &	   $ 48.46\pm  1.67$ \\
ZTF18aaabltn&	 $310.08\pm 77.01$ &	   $177.82\pm  4.61$ &	   $120.02\pm 23.68$ &	   $ 67.66\pm  3.75$ &	   $367.15\pm 33.11$ &	   $312.65\pm  5.71$ &	   $158.14\pm 38.82$ &	   $158.79\pm  4.98$ &	   $ 43.96\pm 11.31$ &	   $ 65.46\pm  2.07$ &	   $ 44.05\pm 11.33$ &	   $ 65.60\pm  2.07$ \\
J0831+3646&	 $119.51\pm 36.89$ &	   $ 97.52\pm  5.53$ &	   $ 35.78\pm 21.51$ &	   $ 29.51\pm  3.58$ &	   $ 45.13\pm 16.60$ &	   $ 86.60\pm  3.55$ &	   $ 41.04\pm 26.62$ &	   $ 65.35\pm  6.70$ &	   $  5.36\pm  9.10$ &	   $ 12.16\pm  3.65$ &	   $  5.37\pm  9.12$ &	   $ 12.18\pm  3.65$ \\
J0909+4747&	 $ 92.32\pm 35.99$ &	   $103.50\pm  6.37$ &	   $ 34.53\pm 27.38$ &	   $ 27.92\pm  4.79$ &	   $171.31\pm 23.31$ &	   $172.98\pm  6.99$ &	   $ 67.51\pm 29.96$ &	   $ 90.66\pm  5.21$ &	   $ 39.29\pm 11.81$ &	   $ 40.93\pm  3.08$ &	   $ 39.38\pm 11.83$ &	   $ 41.02\pm  3.08$ \\
J0937+2602&	 $160.31\pm 91.19$ &	   $132.64\pm  4.17$ &	   $ 45.05\pm 35.96$ &	   $ 33.74\pm  1.72$ &	   $ 69.45\pm 24.97$ &	   $ 68.28\pm  2.40$ &	   $ 93.09\pm 76.94$ &	   $ 86.13\pm  2.59$ &	   $ 32.42\pm 16.55$ &	   $ 22.66\pm  1.28$ &	   $ 32.49\pm 16.58$ &	   $ 22.71\pm  1.29$ \\
J1003+3525&	 $150.32\pm 53.00$ &	   $180.94\pm  4.52$ &	   $107.50\pm 22.34$ &	   $ 57.84\pm  3.59$ &	   $375.77\pm 25.87$ &	   $334.80\pm  8.09$ &	   $ 38.73\pm 48.15$ &	   $105.81\pm  4.26$ &	   $ 16.69\pm 11.23$ &	   $ 51.02\pm  4.16$ &	   $ 16.73\pm 11.26$ &	   $ 51.13\pm  4.17$ \\
J1104+6343&	 $ 47.37\pm  3.10$ &	   $ 46.15\pm  1.92$ &	   $ 19.90\pm  3.69$ &	   $ 13.83\pm  1.80$ &	   $ 36.26\pm  4.44$ &	   $ 37.34\pm  2.60$ &	   $ 20.94\pm  2.97$ &	   $ 21.86\pm  1.53$ &	   $  7.80\pm  2.02$ &	   $ 10.84\pm  1.14$ &	   $  7.82\pm  2.03$ &	   $ 10.86\pm  1.14$ \\
J1110-0003&	 $ 48.03\pm 13.26$ &	   $113.97\pm  5.03$ &	   $ 34.40\pm 12.39$ &	   $ 34.53\pm  3.33$ &	   $113.65\pm 11.10$ &	   $101.98\pm  3.23$ &	   $ 42.21\pm 10.57$ &	   $ 76.85\pm  5.10$ &	   $ 11.23\pm  3.79$ &	   $ 21.07\pm  1.66$ &	   $ 11.26\pm  3.80$ &	   $ 21.11\pm  1.66$ \\
J1115+0544&	 $ 32.30\pm 10.59$ &	   $ 43.05\pm  1.96$ &	   -- &	   $ 14.35\pm  2.89$ &	   $ 62.36\pm 24.36$ &	   $ 20.02\pm  4.38$ &	   $ 11.16\pm 24.09$ &	   $ 47.34\pm  2.11$ &	   $  8.21\pm  5.37$ &	   $ 28.06\pm  2.03$ &	   $  8.22\pm  5.38$ &	   $ 28.12\pm  2.04$ \\
J1132+0357&	 $264.13\pm 28.37$ &	   $198.33\pm  4.52$ &	   $ 69.40\pm 20.22$ &	   $ 44.98\pm  3.35$ &	   $371.57\pm 30.74$ &	   $395.77\pm  9.25$ &	   $293.42\pm 27.00$ &	   $266.19\pm  4.12$ &	   $ 46.61\pm  9.67$ &	   $ 63.59\pm  3.29$ &	   $ 46.71\pm  9.69$ &	   $ 63.73\pm  3.30$ \\
ZTF18aasuray&	 $409.99\pm352.23$ &	   $158.30\pm 12.24$ &	   $ 83.15\pm 59.78$ &	   $ 63.02\pm  3.91$ &	   $292.33\pm 73.71$ &	   $ 95.87\pm  4.91$ &	   $359.36\pm347.99$ &	   $199.08\pm 19.38$ &	   $ 40.63\pm 26.62$ &	   $ 92.28\pm  3.59$ &	   $ 40.71\pm 26.68$ &	   $ 92.47\pm  3.60$ \\
ZTF18aasszwr&	 $152.19\pm 11.46$ &	   $113.58\pm  4.30$ &	   $142.72\pm  8.82$ &	   $ 44.03\pm  5.25$ &	   $132.98\pm  8.83$ &	   $ 84.03\pm  6.17$ &	   $351.68\pm 14.87$ &	   $ 88.05\pm  4.01$ &	   $ 79.68\pm  7.37$ &	   $ 32.83\pm  2.92$ &	   $ 79.85\pm  7.38$ &	   $ 32.90\pm  2.92$ \\
ZTF18aahiqf&	 $174.84\pm 93.32$ &	   $106.56\pm  6.73$ &	   $ 93.45\pm 30.83$ &	   $ 51.88\pm  5.42$ &	   $138.77\pm 42.59$ &	   $ 67.80\pm  9.78$ &	   $158.54\pm 77.53$ &	   $132.49\pm  6.28$ &	   $ 51.49\pm 31.70$ &	   $ 50.44\pm  3.47$ &	   $ 51.60\pm 31.77$ &	   $ 50.55\pm  3.48$ \\
J1259+5515&	 $234.94\pm 76.01$ &	   $104.85\pm  5.69$ &	   $ 33.22\pm 12.96$ &	   $ 29.96\pm  4.15$ &	   $178.37\pm 13.42$ &	   $162.08\pm  3.76$ &	   $171.02\pm 65.82$ &	   $ 91.59\pm  4.65$ &	   $ 38.59\pm  8.71$ &	   $ 28.17\pm  2.61$ &	   $ 38.67\pm  8.73$ &	   $ 28.23\pm  2.62$ \\
J1319+6753&	 $258.14\pm 70.62$ &	   $153.82\pm  4.32$ &	   $157.08\pm 61.46$ &	   $ 55.57\pm  4.31$ &	   $307.28\pm 19.46$ &	   $280.43\pm  5.70$ &	   $290.97\pm 58.03$ &	   $166.11\pm  4.01$ &	   $ 70.84\pm 11.06$ &	   $ 52.77\pm  2.64$ &	   $ 70.99\pm 11.08$ &	   $ 52.89\pm  2.65$ \\
J1358+4934&	 $209.33\pm 21.32$ &	   $191.68\pm  4.54$ &	   $ 73.58\pm  4.80$ &	   $ 59.79\pm  2.98$ &	   $336.28\pm 51.89$ &	   $357.63\pm  9.02$ &	   $ 49.41\pm 10.81$ &	   $ 38.18\pm  3.48$ &	   $ 32.87\pm  3.02$ &	   $ 29.46\pm  1.26$ &	   $ 32.94\pm  3.02$ &	   $ 29.52\pm  1.27$ \\
J1447+2833&	 $387.86\pm  6.07$ &	   $175.14\pm 18.68$ &	   $114.02\pm 11.25$ &	   $ 92.33\pm  3.25$ &	   $299.64\pm 12.48$ &	   $327.11\pm  6.13$ &	   $252.21\pm  5.72$ &	   $298.38\pm  4.51$ &	   $ 63.46\pm  5.38$ &	   $ 79.36\pm  3.26$ &	   $ 63.60\pm  5.39$ &	   $ 79.53\pm  3.26$ \\
ZTF18aajupnt&	 $2543.79\pm128.40$ &	   $177.59\pm  7.94$ &	   $666.31\pm 13.09$ &	   $ 51.20\pm  4.49$ &	   $209.07\pm  5.39$ &	   $ 62.70\pm  9.31$ &	   $592.78\pm 93.09$ &	   $228.71\pm  7.60$ &	   $ 71.15\pm  6.55$ &	   $ 68.92\pm  3.65$ &	   $ 71.30\pm  6.56$ &	   $ 69.07\pm  3.66$ \\
J1533+0110&	 $ 79.54\pm  8.99$ &	   $ 64.50\pm  2.66$ &	   $ 30.68\pm  5.32$ &	   $ 23.35\pm  2.11$ &	   $ 50.54\pm  6.89$ &	   $ 56.08\pm  2.84$ &	   $ 46.98\pm  5.47$ &	   $ 56.25\pm  2.98$ &	   $ 22.45\pm  2.62$ &	   $ 14.91\pm  1.78$ &	   $ 22.50\pm  2.63$ &	   $ 14.94\pm  1.78$ \\
J1545+2511&	 $305.02\pm 17.51$ &	   $274.46\pm  6.96$ &	   $ 60.58\pm  8.27$ &	   $ 66.36\pm  3.77$ &	   $271.67\pm  8.93$ &	   $295.89\pm  6.64$ &	   $223.92\pm 10.81$ &	   $232.23\pm  6.54$ &	   $ 79.20\pm  6.71$ &	   $ 66.75\pm  2.33$ &	   $ 79.37\pm  6.73$ &	   $ 66.90\pm  2.33$ \\
J1550+4139&	 -- &	   $ 49.09\pm  3.65$ &	   $ 10.97\pm 13.84$ &	   $ 21.02\pm  3.41$ &	   $117.72\pm 20.32$ &	   $ 76.27\pm  3.02$ &	   $ 25.46\pm 13.54$ &	   $ 51.63\pm  5.05$ &	   -- &	   $ 13.54\pm  1.56$ &	   -- &	   $ 13.57\pm  1.56$ \\
J1552+2737&	 $ 76.96\pm  5.55$ &	   $ 66.96\pm  3.74$ &	   $ 15.86\pm 10.79$ &	   $ 21.71\pm  2.83$ &	   $112.57\pm 14.49$ &	   $ 67.28\pm  3.78$ &	   $ 56.15\pm  6.38$ &	   $ 71.90\pm  4.69$ &	   $ 26.11\pm  3.73$ &	   $ 32.80\pm  2.43$ &	   $ 26.17\pm  3.74$ &	   $ 32.87\pm  2.43$ \\
J1554+3629&	 -- &	   $ 15.46\pm  5.96$ &	   -- &	   $ 18.98\pm  5.22$ &	   -- &	   $ 58.46\pm  4.88$ &	   -- &	   $ 27.64\pm 10.53$ &	   $ 33.12\pm 14.36$ &	   $  0.00\pm  1.19$ &	   $ 33.19\pm 14.39$ &	   $  0.00\pm  1.19$ \\
PS1-13cbe&	 $333.49\pm  5.51$ &	   $262.57\pm  3.02$ &	   $ 63.82\pm  1.59$ &	   $ 57.39\pm  2.46$ &	   $142.98\pm  1.48$ &	   $111.73\pm  3.37$ &	   $255.20\pm  3.62$ &	   $231.06\pm  2.53$ &	   $ 64.57\pm  0.47$ &	   $ 54.36\pm  1.29$ &	   $ 64.71\pm  0.47$ &	   $ 54.47\pm  1.29$ \\
		
		\hline
	\end{tabular}
\end{sidewaystable*}

\subsubsection{BH mass and Eddington ratio}

With the line profile modeling, we estimate the black hole mass ($\rm M_{BH}$), bolometric luminosity ($\rm L_{bol}$) and Eddington ratio ($\rm L_{bol}/L_{Edd}$), which are critical parameters for describing the AGN activity.

For ``turn-on" state, the $\rm M_{BH}$ of AGNs can be estimated based on the single-epoch spectra with an empirical scaling relation. This relation comes from the assumptions that the BLR is virialized, the monochromatic continuum luminosity is used to estimate the BLR radius, and the broad line width is used as a proxy for the virial velocity. The $\rm {H\beta}$, $\rm {MgII}$, $\rm {CIV}$ and their corresponding continuum luminosities are frequently adopted in such virial calibrations. The $\rm {H\beta}$ is usually used in measuring the $\rm M_{BH}$ for nearby AGNs. However, the emission from $\rm {H\beta}$ is at least a factor of 3 weaker than that of $\rm {H\alpha}$, so, considering about the S/N, $\rm {H\alpha}$ is superior to $\rm {H\beta}$. Using $\rm {H\alpha}$ broad emission line FWHMs and luminosities from the ``turn-on" spectra fitting, we estimate the $\rm M_{BH,vir}$ by following an empirical relation from  \cite{2005ApJ...630..122G}, which is derived by a linear relation between 5100 $\rm \AA$ continuum luminosity and $\rm {H\alpha}$ line luminosity, as well as a strong correlation between the $\rm {H\alpha}$ and $\rm {H\beta}$ emission line widths :

\begin{equation}
M_{BH,vir}=(2.0^{+0.4}_{-0.3})  \qquad \qquad \qquad \qquad \qquad \qquad
\end{equation}

$\times10^{6}(\frac{L_{H\alpha}}{10^{42} ergs s^{-1}})^{0.55\pm0.02}(\frac{FWHM_{H\alpha}}{10^{3} km s^{-1}})^{2.06\pm0.06}M_{\sun}$, 
~\\


~\\

\noindent where $\rm L_{H\alpha}$ is the total $\rm {H\alpha}$ line luminosity. The $\rm L_{bol}$ values in both epochs are estimated from the bolometric correction formula $\rm L_{bol} = 9 \lambda L_{\lambda} (5100$ $\rm \AA)$ \citep{2001ApJ...533..631K}, where $\rm L_{\lambda}$(5100 $\rm \AA)$ is the AGN continuum luminosity at 5100 $\rm \AA$.  
There should be caution in that the estimates of $\rm L_{bol}$ may be uncertain, as the bolometric correction of $\rm L_{\lambda} (5100$ $\rm \AA)$ is based on Type 1 AGNs, and the SED shapes of CL-AGNs are likely different from Type 1 AGNs during their dramatic changes.  The Eddington ratios $\rm L_{bol}/L_{Edd}$ (where $\rm L_{Edd} = 1.26 \times 10^{38}M_{BH,\sigma_{\star}}/M_{\odot}$ $\rm erg$ $\rm s^{-1}$ is the Eddington luminosity)\footnote{The $\rm M_{BH,\sigma_{\star}}$ is measured based on the $\rm M_{BH}-\sigma_{\star}$ relation, which is detailedly described in Section ~\ref{sec:M-sigma}.} are estimated for both the ``turn-on" and ``turn-off" states.

\section{Results and Analysis}

\subsection{$\rm M_{BH}$-$\rm \sigma_{\star}$ relation} 
\label{sec:M-sigma}
\begin{figure}[!htb]
	\begin{center}
		\includegraphics[angle=0,scale=0.36,keepaspectratio=flase]{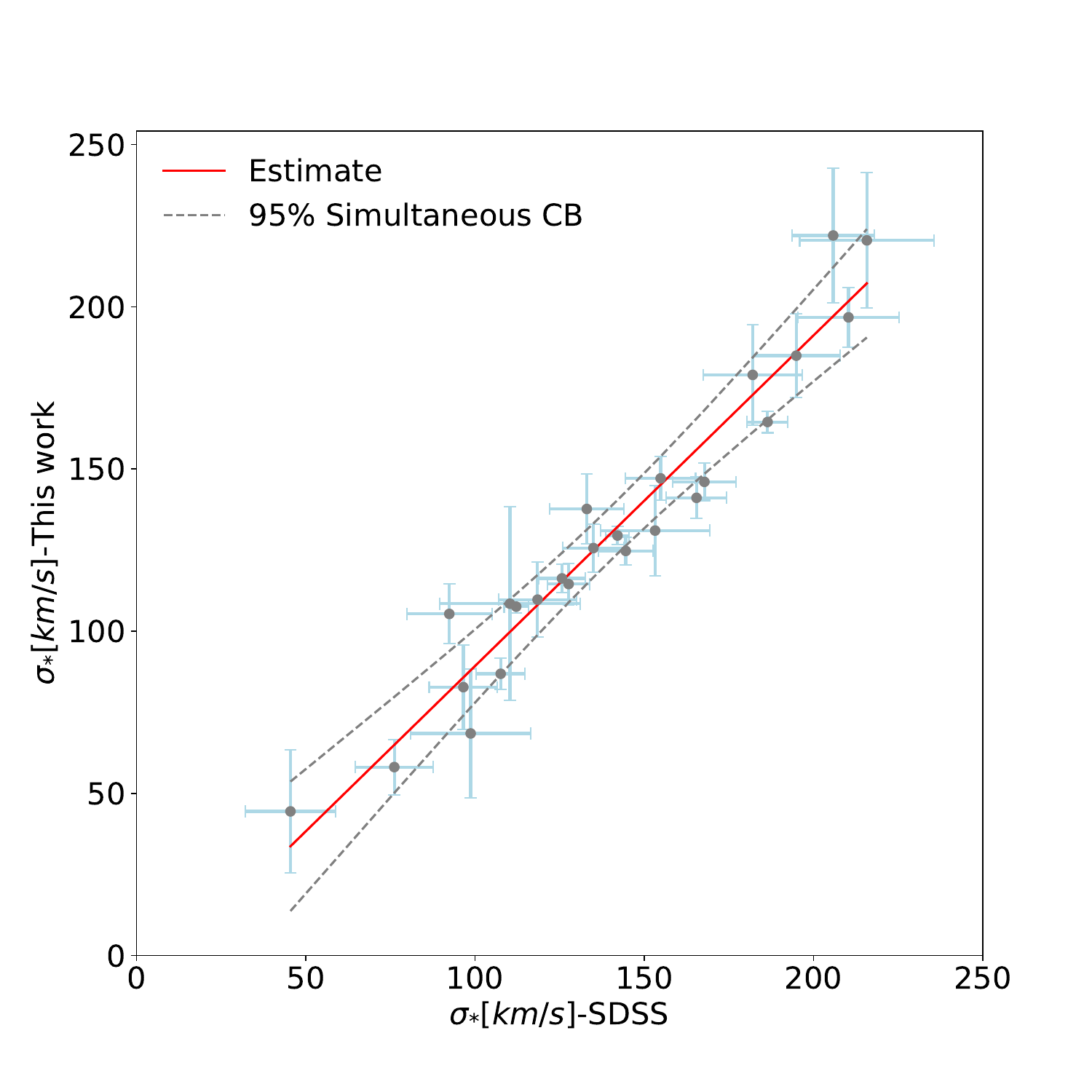}
		\caption{ A comparison of the stellar velocity dispersions estimated by STARLIGHT and those compiled from SDSS. The regression line is traced by the red solid line, and the gray dashed lines indicate the 95\% confidence range using the Scheff$\rm \acute{e}$'s method.  \label{fig:sigma_Com}}
	\end{center}
\end{figure}

The stellar velocity dispersion can be obtained by the fitting of  the ``turn-off" spectra and corrected by the instrumental resolutions of  the SDSS spectra ($\rm \sigma_{inst} \sim 71 km/s$) and base spectra ($\rm \sigma_{base} \sim 70 km/s$). Figure ~\ref{fig:sigma_Com} compares the stellar velocity dispersions obtained from STARLIGHT with the values measured from SDSS. The later ones were measured by the  direct fitting method, which is based on the fact that the individual stellar spectra have been Doppler-shifted because of the star's motion \citep{1961ApJ...133..393B,1992MNRAS.254..389R}.  All objects show a good agreement with a mean and rms difference of  $\rm -9 \pm 12 km$ $\rm s^{-1}$, which makes it accessible for us to investigate the $\rm M_{BH}$-$\rm \sigma_{\star}$ relation of CL-AGNs.

In nearby galaxies, there is apparently a close connection between the central BH and the bulge kinematics. Specifically, BH mass correlates well with stellar velocity dispersion. Many works have shown that AGNs share the same $\rm M_{BH}$-$\rm \sigma_{\star}$ relation as quiescent galaxies \citep{2004ApJ...610..722G,2008ApJ...688..159G}. This relation is essential for understanding the coevolution of BHs and their host galaxies. But some recent works have suggested that this relation may change among different types of host galaxies \citep{2008MNRAS.386.2242H,2009ApJ...698..812G,2009ApJ...698..198G}, different AGN types \citep{2017ApJS..229...39R,2016MNRAS.458L..69B,2008ApJ...678..693M}, or different $\rm M_{BH}$ range \citep{2010ApJ...721...26G}. \cite{2020MNRAS.498.3985Y} and \cite{2021ApJ...907L..21D} suggested that most CL-AGNs in their sample have pseudo-bugle features, and \cite{2014ApJ...789...17H} showed that AGNs with pseudo-bugles have different zero point and much larger scatters than the tight $\rm M_{BH}$-$\rm \sigma_{\star}$ relation for classical bulges.

\begin{figure}[!htb]
	\begin{center}
		\includegraphics[angle=0,scale=0.32,keepaspectratio=flase]{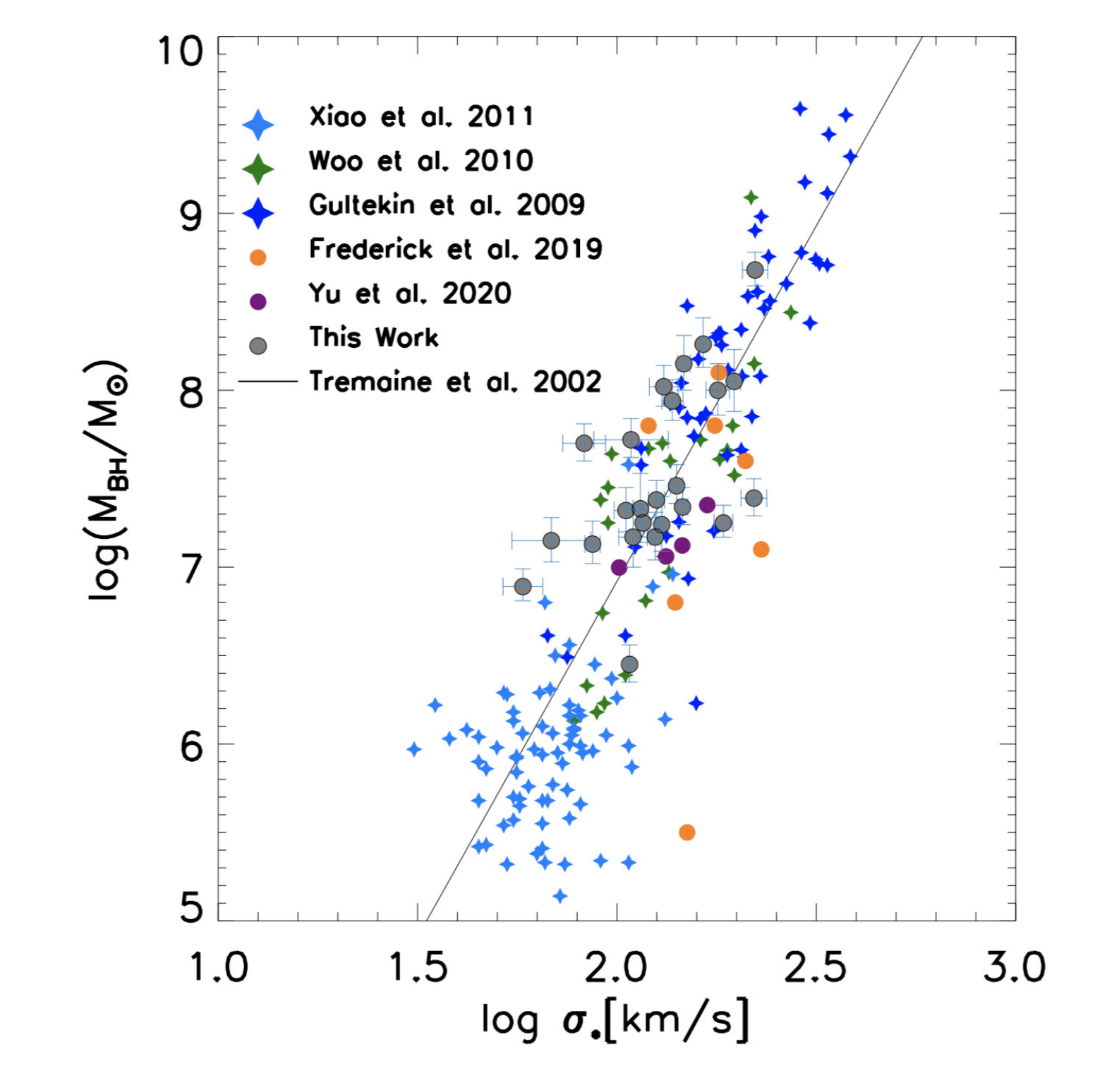}
		\caption{ The $\rm M_{BH}$-$\rm \sigma_{\star}$ relation for CL-AGNs and NCL-AGNs. The filled circles represent CL-AGNs:  gray circles for 25 objects (have both epoch spectra) in our sample; purple circles for 5 Cl-AGNs from \cite{2020MNRAS.498.3985Y} and orange circles for 7 CL-AGNs from \cite{2019ApJ...883...31F}.  The filled stars represent NCL-AGNs: light blue, green and blue stars represent those from \cite{2011ApJ...739...28X}, \cite{2010ApJ...716..269W} and \cite{2009ApJ...698..198G}, respectively. The black solid line is the $\rm M_{BH}$-$\rm \sigma_{\star}$ relation from \cite{2002ApJ...574..740T}. \label{fig:M-sigma}}
	\end{center}
\end{figure}

The CL-AGNs offer a unique opportunity to study this relation. \cite{2017ApJ...835..144G} estimated the $\rm M_{BH,vir}$ of CL-AGN iPTF 16bco through the broad Balmer emission lines when the AGN transformed to Type 1. They reported that the $\rm M_{BH,vir}$ is in good agreement with the $\rm M_{BH,\sigma_{\star}}$ estimated through the $\rm M_{BH}-\sigma_{\star}$ relation. \cite{2020MNRAS.498.3985Y} showed the distribution of 5 CL-AGNs in the $\rm M_{BH}$-$\rm \sigma_{*}$ plane, and found that they follow the $\rm M_{BH}$-$\rm \sigma_{\star}$ relation as NCL-AGNs. \cite{2019MNRAS.485.4790S} used the FWHM of $\rm {H\beta}$ to obtain $\rm M_{BH,vir}$ and their result is in agreement with the estimate based on line dispersion. However, \cite{2019ApJ...883...31F} found that the BH mass inferred from the host galaxy velocity dispersion is larger than the black hole mass estimated from the virial method. \cite{2008ApJ...678..693M} believe that in rapidly accreting objects, the increase in ionizing radiation pressure may lead to the underestimation of $\rm M_{BH,vir}$. The larger size of our sample offers an opportunity to check whether the CL-AGNs also follow the $\rm M_{BH}$-$\rm \sigma_{\star}$ relation. Figure ~\ref{fig:M-sigma} shows the $\rm M_{BH}$-$\rm \sigma_{\star}$ relation for our sample of CL-AGNs. The $\rm M_{BH,vir}$ values are measured from $\rm {H\alpha}$ lines during the ``turn-on" states, and the stellar velocity dispersions are measured from the STARLIGHT fitting results of ``turn-off" spectra.  
As we can see, the CL-AGNs generally follow the same trend as the NCL-AGNs.

While $\rm M_{BH,vir}$ values are more generally used for AGNs, there are large scatters in the $\rm R_{BLR}-L_{5100}$ relation, and it has been established only over a limited range of luminosities; hence, it yields correspondingly uncertain BH masses. On the contrary, the $\rm M_{BH}$-$\rm \sigma_{*}$ relation is a better choice for estimating BH mass. In previous works, direct comparisons have shown that reverberation mapping and stellar velocity dispersions can provide reliable BH mass estimates - within factors of a few, while using the $\rm R_{BLR}-L_{5100}$ relation leads to larger uncertainties \citep{2002ApJ...579..530W}. Therefore, the BH masses that we use in the later sections of this paper are $\rm M_{BH,\sigma_{\star}}$, which are calculated from the $\rm M_{BH}$-$\rm \sigma_{\star}$ relation from \cite{2002ApJ...574..740T}:

\begin{equation}
\begin{split}
log M_{BH,\sigma_{\star}} = (8.13 \pm 0.09) \qquad \qquad \qquad \qquad\\
+(4.04 \pm 0.44) \times log(\sigma_{\star}/200 km s^{-1}). \qquad \quad \qquad
\end{split}
\end{equation}


\subsection{Statistics of the stellar populations in CL-AGN Hosts} 
\label{sec:stellar_population}

\begin{figure*}[!htb]
	\begin{center}
		\includegraphics[angle=0,scale=0.55,keepaspectratio=flase]{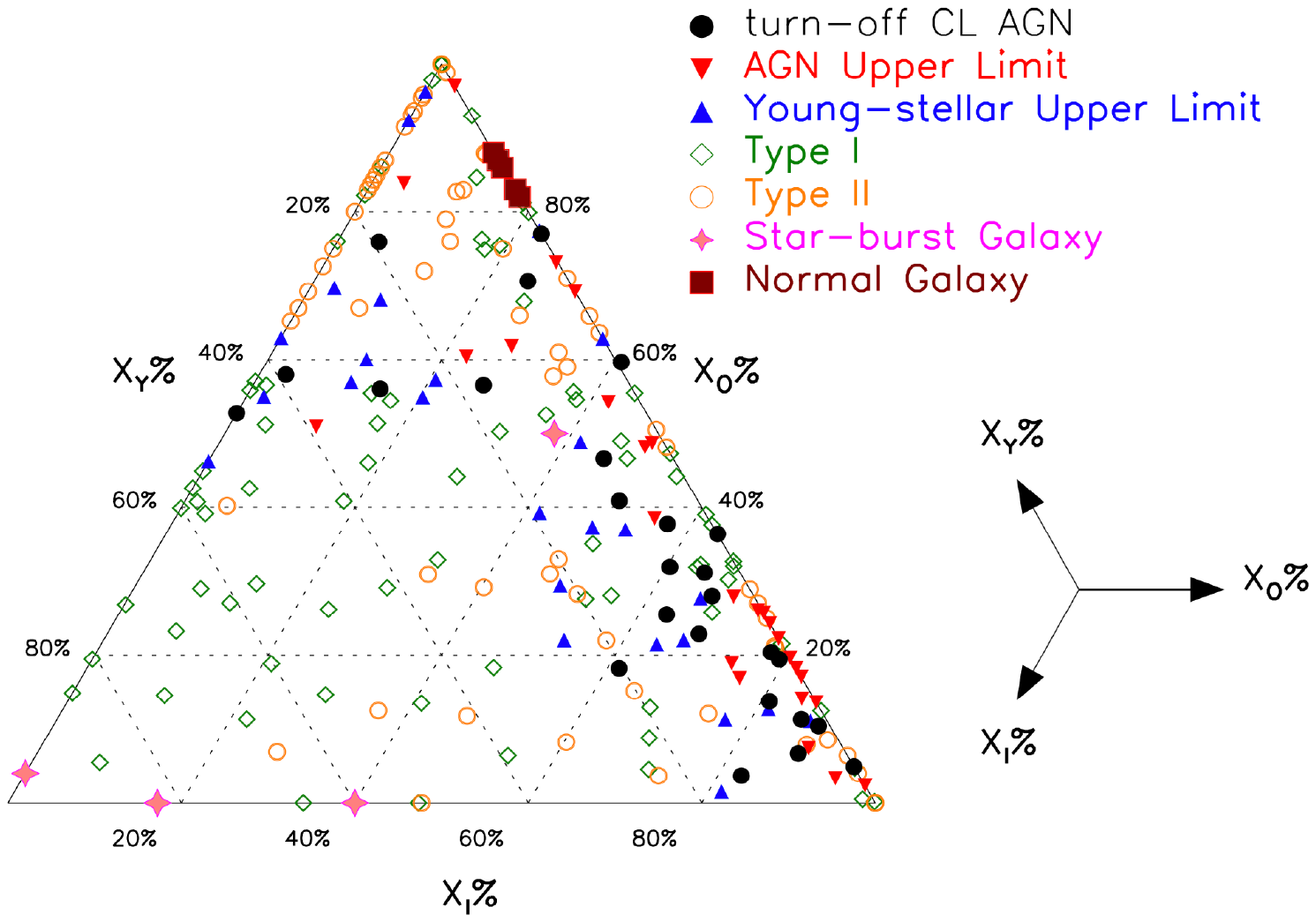}
		\caption{Results of stellar population distributions in a trigonometric coordinate panel. The black filled circles denote the CL-AGNs, the blue filled upward triangles and red filled downward triangles represent the young stellar component upper limit and AGN upper limit (described in Section~\ref{sec:degeneracy}).  Other symbols represent Type 1 AGNs (dark green open diamonds), Type 2 AGNs (orange open circles), star-burst galaxies (pink filled stars) and normal galaxies (dark red filled squares).  The arrows in the bottom right illustrate the directions to $\rm X_{Y}\%$, $\rm X_{I}\%$, $\rm X_{O}\%$.  \label{fig:Stellar_Population}}
	\end{center}
\end{figure*}

In this section, we present the stellar populations in the host galaxies of CL-AGNs. To our knowledge, this is the first time the stellar populations of CL-AGN host galaxies have been presented in detail. Figure ~\ref{fig:Spec_1} has already shown an example of the distributions of the individual stellar populations in a CL-AGN host. However, as suggested by \cite{2004MNRAS.355..273C}, individual stellar population results are uncertain due to the existence of multiple solutions, and the further binning of the ages gives a coarser but more robust description of SFH. Therefore, we bin the stellar populations found by STARLIGHT into three age bins: ``young" ($\rm X_{Y}$,$\rm t \le 0.55\times 10^{8}yr$), ``intermediate" ($\rm X_{I}$,$\rm 1.0\times 1.0^{8}yr\le t \le 1.27\times 10^{9}yr$) and ``old" ($\rm X_{O}$,$\rm t \ge 1.43\times 10^9 yr$). In Figure ~\ref{fig:Stellar_Population}, we present a trigonometric coordinate system in $\rm X_{Y}+X_{I}+X_{O}=1$ panel to show the relative fractions of young, intermediate, and old stellar populations.  As we can see, the CL-AGN hosts are composed of every type of stellar population. Some objects, such as J1554+3629 and J1259+5515, are dominated by old types, with populations older than 1.5 Gyr contributing to $\rm \sim 80\%$ of the emissions at 4020 $\rm \AA$. While in other sources, such as J1358+4934, J1319+6753 and ZTF18aajupnt, the stellar populations are dominated by the intermediate types, with $\rm X_{I} > 85\%$. The young stellar populations are also ubiquitous. They can account for almost half of the emissions in J0159-0052 and ZTF18aaabltn. The stellar populations and other parameters of the CL-AGN host galaxies are summarized in Table~\ref{tab:stellarstellar}.

\begin{table*}
	\setlength{\tabcolsep}{10pt}
	\caption{The stellar populations and black hole masses ($\rm M_{BH,\sigma_{\star}}$) of 26 CL-AGNs. 	\label{tab:stellarstellar}}
	\begin{tabular}{lccccccc} 
	\hline
	\hline
	
		name  & $\rm x_Y$     & $\rm x_I$     & $\rm x_O$     & $\rm x_{FC}$ & $\rm <logt_{\star}>_{L}$      & $\rm \sigma_{\star}$ & $\rm M_{BH,\sigma_{\star}}$ \\
		& $\rm (\%)$     & $\rm (\%)$     & $\rm (\%)$     & $\rm (\%)$ & log(yr)     & $\rm km$ $\rm s^{-1}$ & $\rm log(M/M_{\odot})$ \\
		\\
		(1)        & (2)     & (3)     & (4)   & (5) &(6)  & (7)    & (8)    \\
	\hline
	J0126-0839   & $17.0\pm3.3$  &  $26.7\pm  3.5$  &  $56.5\pm  2.1$  &  $10.4\pm  2.1$  &  $ 9.62\pm 0.04$  &$ 82\pm 13$  &  $ 6.59\pm 0.34$ \\
	J0158-0052   & $47.2\pm2.7$  &  --  &  $52.8\pm  2.5$  &  $16.8\pm  4.8$  &  $ 9.47\pm 0.09$  &$ 44\pm 19$  &  $ 5.50\pm 0.81$ \\
	J0159+0033   & --  &  $95.1\pm  3.9$  &  $ 4.9\pm  3.9$  &  $22.0\pm  1.5$  &  $ 8.86\pm 0.05$  &$ 118\pm 26$  &  $ 7.22\pm 0.41$ \\
	ZTF18aaabltn   & $39.3\pm4.4$  &  $ 3.4\pm  3.8$  &  $58.0\pm  3.5$  &  $ 6.8\pm  3.9$  &  $ 9.61\pm 0.06$  &$ 86\pm 4$  &  $ 6.67\pm 0.21$ \\
	J0831+3646   & $ 9.2\pm3.1$  &  $50.2\pm 10.2$  &  $40.9\pm 10.3$  &  $ 7.4\pm  3.1$  &  $ 9.24\pm 0.04$  &$ 196\pm 9$  &  $ 8.10\pm 0.12$ \\
	J0909+4747   & --  &  $63.6\pm 13.8$  &  $36.4\pm 13.8$  &  $14.7\pm  1.2$  &  $ 9.08\pm 0.02$  &$ 105\pm 9$  &  $ 7.01\pm 0.22$ \\
	J0937+2602   & $ 8.7\pm2.8$  &  $61.4\pm  6.9$  &  $31.9\pm  6.5$  &  $15.5\pm  2.8$  &  $ 9.20\pm 0.04$  &$ 114\pm 6$  &  $ 7.16\pm 0.17$ \\
	J1003+3525   & $ 4.5\pm1.9$  &  $65.1\pm  8.4$  &  $31.2\pm  8.2$  &  $12.7\pm  2.4$  &  $ 9.24\pm 0.03$  &$ 147\pm 6$  &  $ 7.59\pm 0.13$ \\
	J1104+6343   & $ 4.9\pm6.0$  &  $24.8\pm  8.8$  &  $70.6\pm  9.6$  &  $16.7\pm  5.2$  &  $ 9.38\pm 0.09$  &$ 108\pm 29$  &  $ 7.06\pm 0.50$ \\
	J1110-0003   & $ 8.8\pm4.9$  &  $68.2\pm  8.8$  &  $22.9\pm  9.3$  &  $22.8\pm  5.1$  &  $ 9.24\pm 0.09$  &$ 131\pm 13$  &  $ 7.39\pm 0.22$ \\
	J1115+0544   & $ 8.1\pm1.4$  &  $45.5\pm  7.0$  &  $46.6\pm  7.4$  &  $ 2.0\pm  1.4$  &  $ 9.24\pm 0.04$  &$ 116\pm 4$  &  $ 7.18\pm 0.15$ \\
	J1132+0357   & $ 5.0\pm1.7$  &  $57.1\pm  5.5$  &  $37.8\pm  5.6$  &  $ 5.9\pm  1.5$  &  $ 9.35\pm 0.02$  &$ 124\pm 4$  &  $ 7.31\pm 0.14$ \\
	ZTF18aasuray   & $11.4\pm1.8$  &  $63.2\pm  5.5$  &  $25.5\pm  5.7$  &  $ 5.2\pm  2.2$  &  $ 9.30\pm 0.04$  &$ 129\pm 2$  &  $ 7.37\pm 0.13$ \\
	ZTF18aasszwr   & $ 0.9\pm1.9$  &  $78.7\pm  8.2$  &  $19.4\pm  8.2$  &  --  &  $ 9.22\pm 0.05$  &$ 221\pm 20$  &  $ 8.31\pm 0.19$ \\
	ZTF18aahiqf   & --  &  $41.4\pm  5.8$  &  $59.7\pm  5.8$  &  --  &  $ 9.46\pm 0.03$  &$ 164\pm 3$  &  $ 7.79\pm 0.10$ \\
	J1259+5515   & $19.0\pm3.5$  &  $ 4.5\pm  5.5$  &  $75.9\pm  6.9$  &  $ 6.1\pm  4.9$  &  $ 9.52\pm 0.05$  &$ 220\pm 20$  &  $ 8.30\pm 0.19$ \\
	J1319+6753   & $ 5.2\pm2.6$  &  $87.4\pm  4.7$  &  $ 6.7\pm  4.2$  &  $11.7\pm  2.2$  &  $ 8.93\pm 0.01$  &$ 109\pm 11$  &  $ 7.08\pm 0.24$ \\
	J1358+4934   & --  &  $86.9\pm  5.7$  &  $10.4\pm  5.9$  &  $29.2\pm  3.3$  &  $ 9.09\pm 0.03$  &$ 68\pm 19$  &  $ 6.26\pm 0.56$ \\
	J1447+2833   & $13.1\pm2.3$  &  $82.3\pm  3.9$  &  $ 3.7\pm  3.7$  &  $13.2\pm  2.0$  &  $ 8.86\pm 0.02$  &$ 141\pm 6$  &  $ 7.52\pm 0.14$ \\
	ZTF18aajupnt   & $ 3.3\pm1.5$  &  $86.2\pm  3.2$  &  $11.3\pm  3.2$  &  $ 3.0\pm  1.5$  &  $ 9.22\pm 0.03$  &$ 107\pm 2$  &  $ 7.05\pm 0.15$ \\
	J1533+0110   & $ 2.2\pm1.5$  &  $78.2\pm  7.1$  &  $20.4\pm  6.8$  &  $ 7.1\pm  1.8$  &  $ 9.31\pm 0.04$  &$ 125\pm 7$  &  $ 7.32\pm 0.16$ \\
	J1545+2511   & $20.7\pm2.9$  &  $61.6\pm  5.2$  &  $18.2\pm  4.9$  &  $ 9.6\pm  2.7$  &  $ 9.11\pm 0.05$  &$ 146\pm 5$  &  $ 7.58\pm 0.13$ \\
	J1550+4139   & $ 4.9\pm2.6$  &  $80.5\pm  4.0$  &  $13.8\pm  3.8$  &  $ 6.7\pm  2.2$  &  $ 9.30\pm 0.04$  &$ 184\pm 12$  &  $ 7.99\pm 0.15$ \\
	J1552+2737   & $ 5.1\pm1.9$  &  $67.5\pm  6.3$  &  $28.0\pm  6.8$  &  $10.0\pm  2.1$  &  $ 9.36\pm 0.06$  &$ 137\pm 10$  &  $ 7.48\pm 0.18$ \\
	J1554+3629   & --  &  $23.0\pm  3.7$  &  $77.0\pm  3.7$  &  --  &  $ 9.83\pm 0.01$  &$ 178\pm 15$  &  $ 7.94\pm 0.18$ \\
	PS1-13cbe   & $29.2\pm3.7$  &  $15.0\pm  6.8$  &  $56.0\pm  5.6$  &  $ 0.9\pm  2.2$  &  $ 9.25\pm 0.04$  &$ 58\pm 8$  &  $ 5.97\pm 0.36$ \\
\hline
\end{tabular}
	\tablecomments{Columns 2-5 list the percentage light fractions associated to young components ($\rm x_Y$), intermediate components ($\rm x_I$), old components ($\rm x_O$) and AGN featureless continuum ($\rm x_{FC}$), respectively. Column 6 lists the mean age of CL-AGN hosts weighted by light .Column 7 lists the velocity dispersion derived from STARLIGHT fitting. Column 8 lists the black hole mass calculated from the $\rm M_{BH}$-$\rm \sigma_{\star}$ relation.}
\end{table*}

\begin{figure}[!htb]
	\begin{center}
		\includegraphics[angle=0,scale=0.45,keepaspectratio=flase]{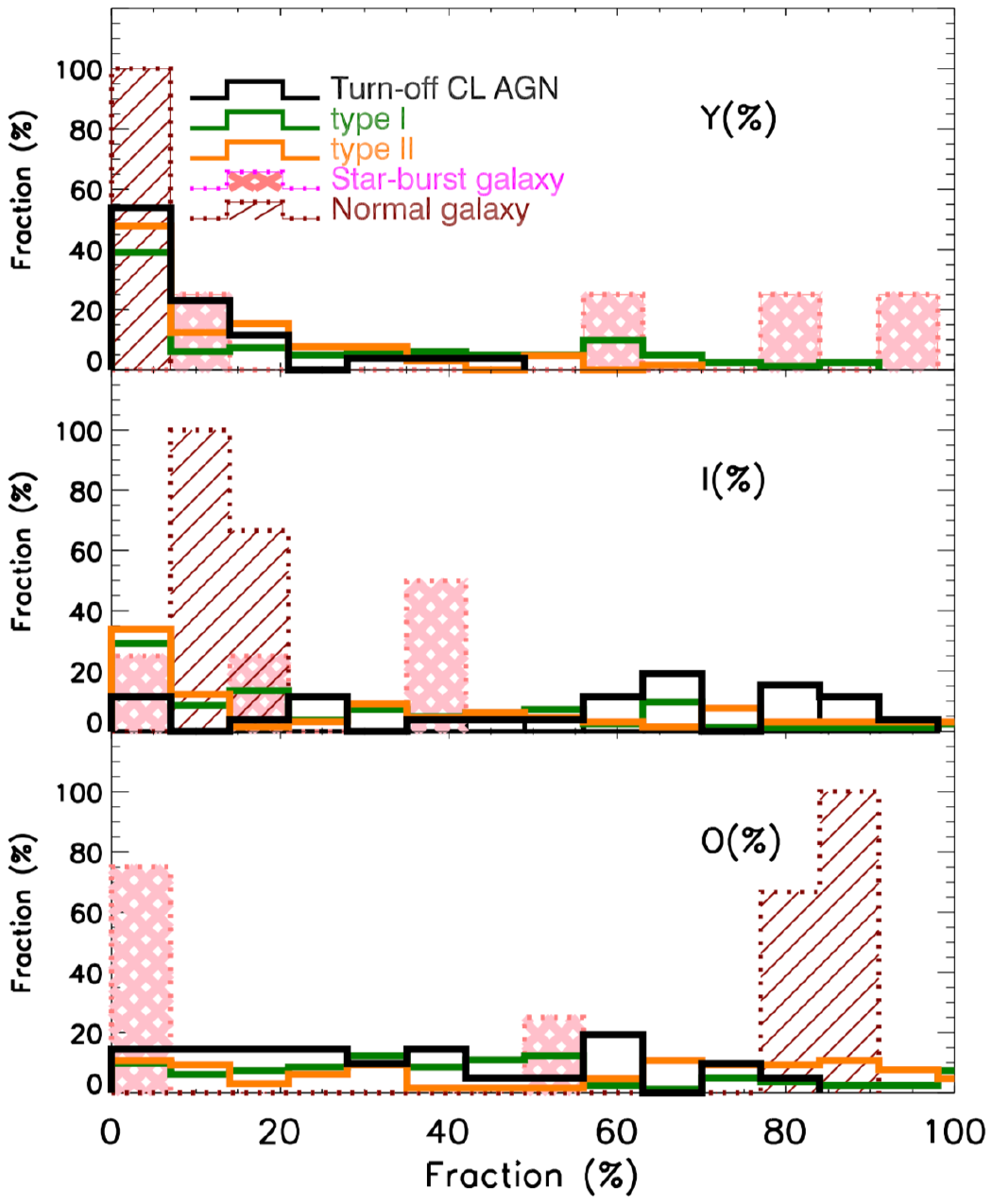}
		\caption{The same as Figure ~\ref{fig:Stellar_Population}, but depicted in the form of histograms.\label{fig:Stellar_Population2}}
	\end{center}
\end{figure}

We compare our results with those reported by  \cite{2004MNRAS.355..273C} and \cite{2018ApJ...864...32J} for other types of AGN host galaxies. The methods they used also followed a description of stellar populations plus a power-law component, similar to us. The stellar populations in these three samples can be compared conveniently and accurately. The sample in \cite{2004MNRAS.355..273C} contains 65 Seyfert 2s, 4 starburst galaxies (SBGs), and  5 normal galaxies \footnote{Given the redshift range of objects in \cite{2004MNRAS.355..273C} is lower than that of CL-AGNs in our sample, the comparisons may be affected by the fact that SFR is evolved with redshift. However, we compared the stellar distributions in high-z end of Seyfert 2s to that of whole Seyfert 2 sample, as well as the stellar distributions in the low-z end of CL-AGNs to that of the entire CL-AGNs sample. We found that the distributions of stellar populations in different redshift ranges show no visible differences, so we believe that the results of comparisons are reliable.}. The spectra were observed by the 1.5m ESO telescope at La Silla (Chile), with a wavelength range from 3470 to 5450 $\rm \AA$ and a spatial sampling of $\rm 0.83^{\prime\prime}$ per pixel. In \cite{2018ApJ...864...32J}, there are in total 82 Type 1 QSOs with SDSS spectra.  Figure ~\ref{fig:Stellar_Population} and Figure ~\ref{fig:Stellar_Population2} (the same result as Figure ~\ref{fig:Stellar_Population}, but in the form of histograms) summarize the results of comparisons. The stellar populations of CL-AGN and NCL-AGN hosts seem to be more homogeneous when compared with SBGs and normal galaxies. The AGN hosts (CL-AGNs, normal Type 1s and Type 2s) lie between SBGs and normal galaxies. All five normal galaxies are dominated by old stellar populations ($\rm X_{O} \gtrsim 80\%$), while most SBGs are dominated by the young stellar populations. This is consistent with previous works which showed no significant differences between CL-AGN and NCL-AGN hosts \citep{2019ApJ...876...75C,2020MNRAS.498.3985Y,2021ApJ...907L..21D,2021ApJ...915...63L}, and both primarily reside in the region between star-forming galaxies and normal galaxies. However, when the detailed stellar populations are compared among AGNs with different spectral types, it shows a little difference (more clear in the histogram of  Figure ~\ref{fig:Stellar_Population2}): Type 1 AGN hosts have a long tail toward high $\rm Y(\%)$, while the old stellar populations contribute a few more in Type 2 AGN hosts. For the CL-AGN hosts, there are more significant contributions from the intermediate components, and the fraction of young components is relatively small.

We perform the Kolmogorov-Simirov (KS) tests on the stellar populations of any two AGN samples, and the yielded probabilities $\rm P_{KS}$ are given in Table~\ref{table:KS}. It can be seen that the $\rm P_{KS}$ values of any two comparisons (except the comparison of old stellar populations between Type 1 and Type 2 AGN hosts) are greater than 0.001,  which indicates that there are no significant differences in the stellar populations among different kinds of AGN hosts. On the other hand, there are more contributions from young stellar populations in the host galaxies of Type 1 AGNs. This difference is relatively significant, with $\rm P_{KS} \le 0.05$ in the comparison of type 1 vs. type 2 and type 1 vs. CL-AGN in young stellar population. Similarly, $\rm P_{KS}$ values are also less than 0.05 in the comparisons of type 2 vs. type 1 and type 2 vs. CL-AGN in old stellar populations, as well as CL-AGN vs. type 1 and CL-AGN vs. type 2 in intermediate stellar populations. This is consistent with our previous description.

\begin{table}[h]
	\caption{The probability in the KS-test for stellar population. \label{table:KS}}
	\begin{tabular}{c}
		\begin{minipage}[b]{1.\columnwidth}
			\centering
			\raisebox{-.5\height}{\includegraphics[width=\linewidth]{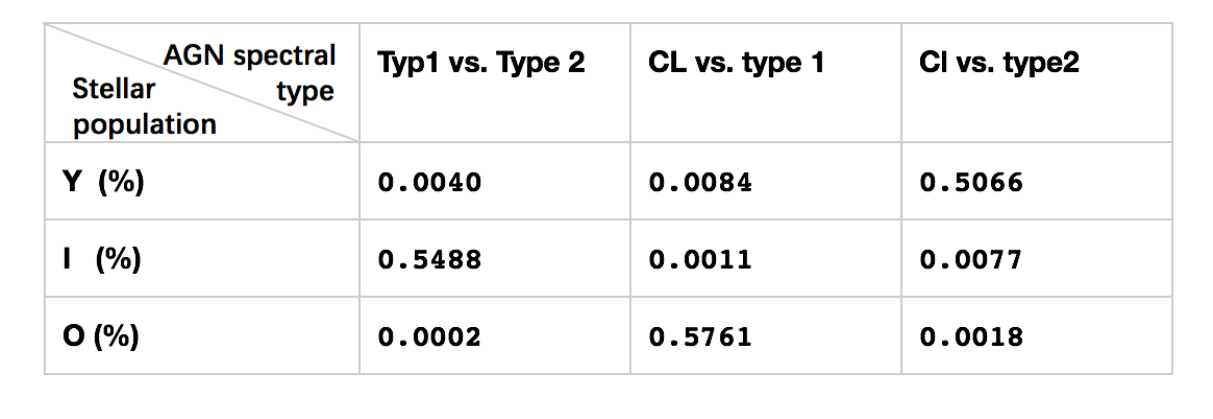}}
		\end{minipage}
	\end{tabular}
	\tablecomments{The numbers show the probability $\rm P_{KS}$ between different  AGN spectral types.} 
\end{table}

\subsection{Host Galaxy Classifications } \label{sec:BPT}

\begin{figure*}[!htb]
	\begin{center}	
		\includegraphics[page=1,scale=0.6]{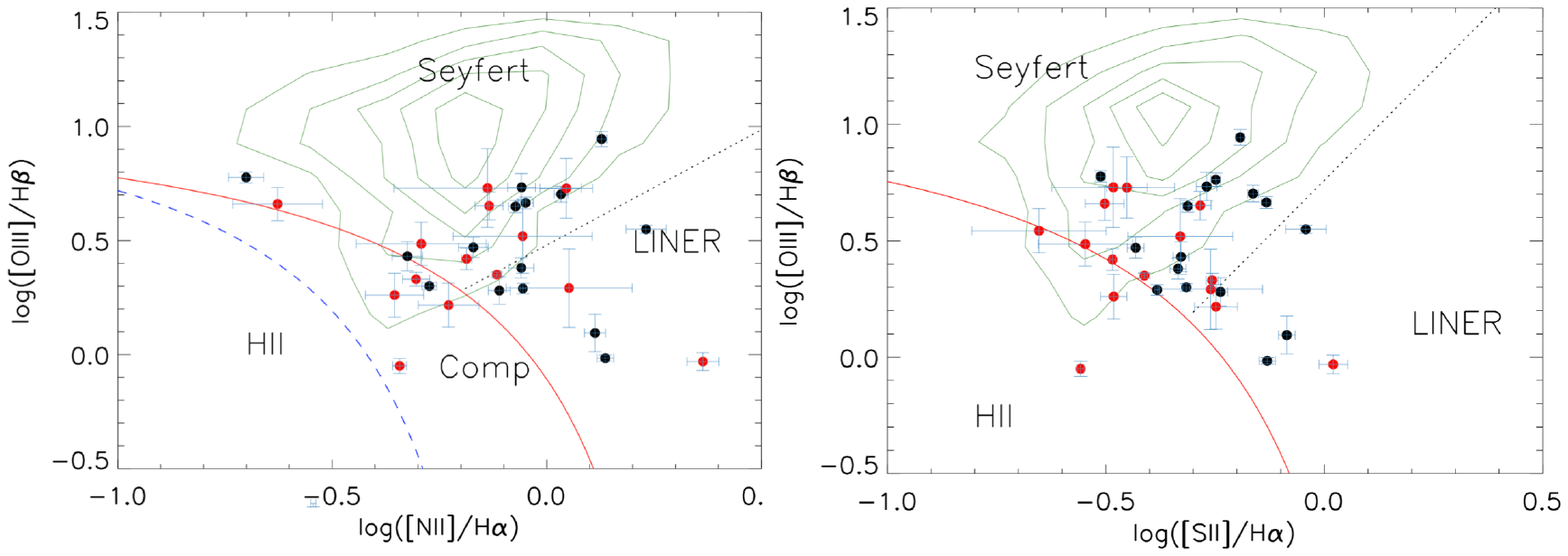}   
		\caption{The BPT diagrams using  $\rm {[NII]/H\alpha}$, $\rm {[OIII]/H\beta}$ and $\rm {[SII]/H{\alpha}}$. The lines represent the classification lines for the pure star-formation (blue dashed line, \citealt{2003MNRAS.346.1055K}), the extreme star-formation (red sold lines, \citealt{2001ApJ...556..121K}) and the Seyfert-LINER classification (black dot lines, \citealt{2006MNRAS.372..961K,2010MNRAS.403.1036C}). The off- and on-state are represented by black and  red dots,  respectively. The dark green contours indicate the distributions  of SDSS quasars with $\rm z < 0.35$ from  \cite{2011ApJS..194...45S}.}
	\end{center}
	\label{fig:BPT} 
\end{figure*}

The $\rm {[NII]/H{\alpha}}$ versus $\rm {[OIII]/H{\beta}}$ diagnostic diagram (the "Baldwin, Phillips \& Terlevich", BPT diagram) is commonly used to distinguish AGN activity from star formation \citep{1981PASP...93....5B,1987ApJS...63..295V}.  In Figure ~\ref{fig:BPT} we show the line ratios of CL-AGNs in the BPT diagram. The CL-AGNs with unresolved narrow lines are excluded (the $\rm S/N$ of narrow lines are required to be greater than 2.0). We also show the pure star formation line \citep{2003MNRAS.346.1055K}, the extreme star formation line \citep{2001ApJ...556..121K}, and the Seyfert-LINER classification line  \citep{2006MNRAS.372..961K,2010MNRAS.403.1036C}. The emission line ratios of the ``turn-off" spectra, after the subtraction of the stellar components (both continuum and emission lines), classify most CL-AGN host galaxies into a borderline region between the LINER and Seyfert classifications, and some of others are consistent with AGN-like, or composite (between AGN and HII) regions, which are clearly different from those of normal AGNs. This result is consistent with previous works. Three CL-AGNs in \cite{2016ApJ...826..188R} were classified as AGN-like. The emission line ratio of iPTF 16bco is consistent with a LINER \citep{2017ApJ...835..144G}.  PS1-13cbe is classified as a LINER in the automatic Portsmouth pipeline from SDSS classification. However, calculations after starlight subtraction classify it clearly as a Seyfert \citep{2018arXiv181103694K}. Five newly-found CL-AGNs in \cite{2019ApJ...883...31F} are also located in the region between LINER and Seyfert. 

We also present the emission line ratios for the ``turn-on" spectra in Figure ~\ref{fig:BPT}. It is clear that ``turn-on" states have similar distributions as ``turn-off" states. This similar behavior between two phases may be explained by the fact that the narrow line emissions trace the average AGN continuum over a much longer timescale than changing-look transition. While the emissivity decay time for narrow lines is only a few years, the light-travel time effects prolong the response time to hundreds of years. Therefore, any short-term variabilities are smeared out \citep{1995ApJ...445L...1E}. So it may not be surprising that the ``turn-on" and ``turn-off" states have similarly narrow emission line ratios.

It should be noted that the spectroscopic apertures of ``turn-off" spectra are relatively large and they can encompass the emissions from host galaxies \citep{2002ApJ...579L..71M,2014MNRAS.441.2296M,2019ApJ...876...12A}. Even with the stellar components subtraction, the emission-line ratios can still be affected by the aperture, as the emission lines are enhanced by the extranuclear SF activities \citep{2014MNRAS.441.2296M}. What's more, the aperture sizes used in the on- and off-state observations of the same object may be different, which will bring additional uncertainties. However, the spectra of the control sample (the normal AGNs) in Figure ~\ref{fig:BPT} were also observed by SDSS, so we believe that the comparison between ``turn-off" CL-AGNs and normal AGNs is meaningful. As for the cases of ``turn-on" CL-AGNs, the overall distribution of emission-line ratios does not change when the aperture effects are considered\footnote{Assuming that the emissions from the host galaxies detected and measured in the spectroscopic aperture are uniformly distributed, the starlight in the on-state aperture are corrected by the factor $\rm A_{on}/A_{off}$, which is the aperture size ratio between on- and off-observations. }, so we still keep the BPT diagram results of the ``turn-on" CL-AGNs as a reference. 

\subsection{Distribution in BH parameter space} 
\label{sec:BH_para}

\begin{figure*}[!htb]
	\begin{center}	
		\includegraphics[page=1,scale=0.5]{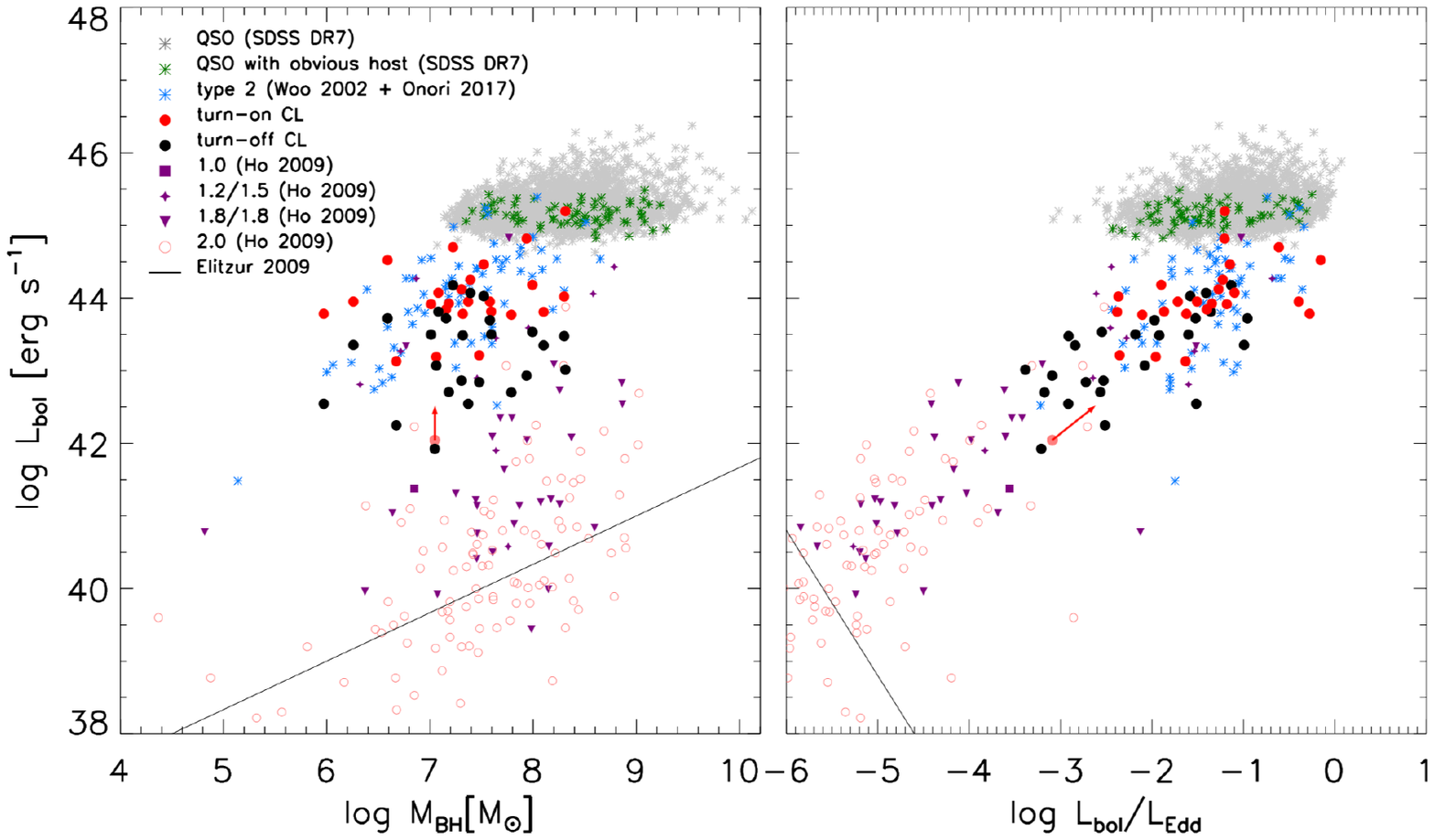}   
		\caption{The parameter space of $\rm M_{BH}$, $\rm L_{bol}$ and $\rm L_{bol}/L_{Edd}$. The light gray crosses denote the SDSS quasar with $\rm z < 0.35$ \citep{2011ApJS..194...45S}, and those have obvious stellar components \citep{2018ApJ...864...32J} are represented by dark green crosses. The blue crosses represent Type 2 AGNs from \cite{2002ApJ...579..530W}. The low-luminosity ``dwarf" Seyfert sample from \cite{2009ApJ...701L..91E} are represented by purple symbols (intermediate type) and red open circles (Type 2). We overplot the CL-AGNs sample in their on (red filled circles) and off (black filled circles) states.  The red arrow shows the upper limit (assuming no contribution from young component) of ZTF18aajupnt. For other objects, there are no much differences between this extreme assumption and normal fitting (see description in Section~\ref{sec:degeneracy}), so we don't mark the upper limits for other them. The black solid lines represent the critical value below which the BLR disappears \citep{2009ApJ...701L..91E}.
		}
	\end{center}
	\label{fig:BH_para} 
\end{figure*}

When AGNs are separated into different types by spectral classification, they show a rough evolutionary sequence in BH parameter space (BH mass $\rm M_{BH}$, bolometric luminosity $\rm L_{bol}$ and Eddington ratio $\rm L_{bol}/L_{Edd}$).  Figure ~\ref{fig:BH_para} shows the distributions of AGNs with different spectral classifications, which occupy distinct regions in the BH parameter space. When CL-AGNs ``turn-on", their distributions roughly lie between Type 1 and Type 2 AGNs. When they ``turn-off", they largely overlap with Type 2 AGNs, and some of them even extend to the region of ``dwarf" Seyfert galaxies. Many recent works have pointed out that CL-AGNs are biased toward low-luminosity and low Eddtington ratio \citep{2019ApJ...883...31F,2019ApJ...887...15W,2019ApJ...874....8M,2020MNRAS.498.3985Y,2020ApJ...888...58G,2020ApJ...905...52G}. Combining this bias and the AGN evolutionary sequence in the BH parameter space, \cite{2019ApJ...874....8M} and \cite{2019ApJ...883...31F} have proposed that the disk-wind model is plausible for the CL phenomena of CL-AGNs. 
We will give a detailed discussion of it in Section~\ref{sec:BH_para_discussion}

ZTF18aajupnt is a narrow-line Seyfert 1 (NLS1) during its ``turn-on" state, and its position in the BH parameter space deviates from other CL-AGNs with lower $\rm L_{bol}$  and $\rm L_{bol}/L_{Edd}$ values. Figure ~\ref{fig:BH_para} shows that the upper limit values (the extreme assumption of no contribution from the young components) of ZTF18aajupnt fall into the region of other CL-AGNs. ZTF18aajupnt was initially classified as a TDE because of its Balmer and HeII emission features in the optical spectrum. However,  \cite{2019ApJ...883...31F} showed that its ZTF light curve together with the UV spectrum and X-ray monitoring are more consistent with CL-AGNs. So we still include ZTF18aajupnt in our CL-AGN sample. Another possible reason for this deviation is that the estimated projected angular size of the NLR is larger than the aperture size during the on-state observation. When considering the aperture correction of the cross-calibration, ZTF18aajupnt also falls into the region of other CL-AGNs. The combined effects of the specific features in the optical spectrum and the observational aperture size may be both be reasons for its deviation in the BH parameter space.

\section{Discussion}

\subsection{Nature of the Changing state}
\label{sec:BH_para_discussion}

As mentioned in Section ~\ref{sec:BPT}, CL-AGNs are classified into a borderline region in the BPT diagram between LINER and Seyfert. Some works \citep{2006ApJ...653L..25G,2009ApJ...698.1367G,2012Natur.485..217G,2014ApJ...780...44C,2014MNRAS.445.3263H,2016ApJ...826..188R} have shown that the emission line ratios of TDEs are consistent with star-formation rather than AGN. This can also be explained by the long light travel time for  TDE continuum emissions to reach the NLR. However, some studies have suggested that TDEs can happen in AGN host galaxies, given that the dense star-formation clouds and preexisting accretion disk can increase the chance of stars encountering tidal disruption and reduce the relaxation time \citep{2006JPhCS..54..293P,2017ApJ...843..106B}. In fact, some TDEs were indeed found in weak AGNs \citep{2011ApJ...740...85W,2017ApJ...843..106B}.  So it's hard to exclude the TDE scenario based on the CL-AGN BPT diagram alone. Further works on  comparing the distributions of the CL-AGNs and the confirmed TDEs on the BPT diagram will be helpful for solving this problem.

\begin{figure*}[!htb]
	\begin{center}	
		\includegraphics[page=1,scale=0.7]{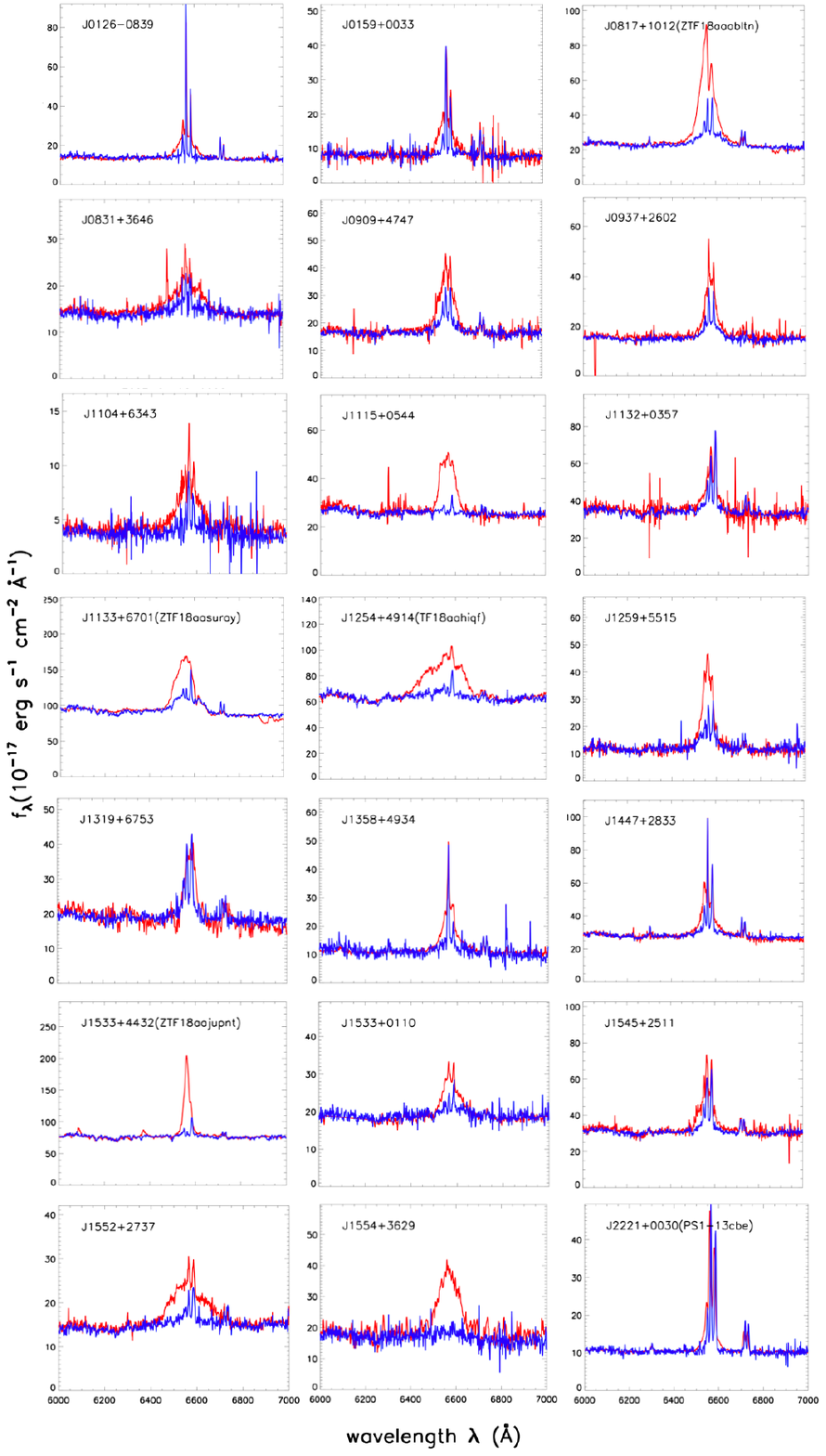}   
		\caption{The 21 decomposed AGN spectra in the $\rm H\alpha$ region. The ``turn-on" spectra (red) are compared to the dereddened ``turn-off" spectra (blue).  Although the dimming in continuum emissions can be explained by the changes in dust extinction, in most case, the predicted strength of the broad $\rm H\alpha$ components is much weaker than expected, which indicates that extinction origin alone can not explain the extraordinary changes in CL-AGNs. The objects J1003+3525,  J1110-0003, J1225+5108, J1550+4139 are not included in this figure due to the limited wavelength range of continuum in the $\rm H\alpha$ region or low spectral resolution.}
	\end{center}
	\label{fig:obscura} 
\end{figure*}

Besides TDEs, there is still an on-going argument whether the CL phenomenon is intrinsic to the changes of the central engine, or just a result of gas clouds moving in and out along the line of sight. If we assume that the CL phenomenon  is cased by the intervening material along the line of sight outside the BLR, there will be no significant differences between decomposed ``turn-on'' AGN spectra and dereddened, decomposed ``turn-off" AGN spectra. Following the methods of \cite{2015ApJ...800..144L} and \cite{2016ApJ...826..188R}, a reddening law \citep{1989ApJ...345..245C} for the dust extinction with $\rm R_{v} = 3.1$ is applied to decompose the AGN spectra from the ``turn-off" state in the $\rm H\alpha$ region, allowing E(B-V) to be free. The best fitted value of E(B-V) is determined by minimizing the $\rm \chi^2$ between the ``turn-on" and ``turn-off" continuum emissions within the $\rm H\alpha$ region. Figure ~\ref{fig:obscura} compares the fitted dereddened ``turn-off" spectra in the $\rm H\alpha$ region with the ``turn-on" spectra. For the object J1319+6753, when considering the low spectral resolution of  the ``turn-on" state, the emission line profile of the dereddend ``turn-off" spectrum is similar to that of the ``turn-on" spectrum. But for other objects, it is clear that the $\rm H\alpha$ emission complex in ``turn-off" states can not be reproduced by the same extinction model, which is similar to the results of \cite{2015ApJ...800..144L} and \cite{2016ApJ...826..188R}. The larger sample size in our work further suggests that the dust extinction scenario alone can not be responsible for the changes in the CL-AGN spectra. As for the object J1319+6753, observations with higher spectral resolution are required to check whether the changes in the optical spectra of J1319+6753 can indeed be modeled by dust extinction.

As mentioned in Section~\ref{sec:BH_para}, the $\rm L_{bol}$ and $\rm L_{bol}/L_{Edd}$ values of our CL-AGN sample are biased toward lower values.  Especially for objects in the``turn-off" state, the faint end has approached to the region of the low-luminosity “dwarf” Seyfert 2. Some recent works argue that many of these low-luminosity AGNs are ``true" Type 2 AGNs, which intrinsically lack BLR due to inadequate accretion rate \citep{2001ApJ...554L..19T,2003ApJ...583..632T,2003ApJ...590...86L,2008ARA&A..46..475H}. The existence of such kind of sources implies that the AGN spectral type can be changed due to the accretion rate changes. The emission line ratio distributions of CL-AGNs on the BPT diagram also indicate that they lack strong accretion signatures.  All of these results are consistent with the well-known anti-correlation between $\rm L_{bol}/L_{Edd}$ and variability amplitude  for quasars \citep{2008MNRAS.383.1232W,2009RAA.....9..529M,2010ApJ...721.1014M,2014MNRAS.444.3078G} and suggest that CL-AGNs may be associated with the accretion rate change of the central BH.

Similar to the results of \cite{2019ApJ...874....8M} and \cite{2019ApJ...883...31F},when compared to the normal AGNs, the $\rm L_{bol}$ and $\rm L_{bol}/L_{Edd}$  values of CL-AGNs during the ``turn-off" states are closer to the critical values below which the BLR disappears. These critical values are derived from the disk-wind model \citep{2014MNRAS.438.3340E}, where the quenching of BLR at a low accretion rate is an unavoidable consequence and the AGNs follow an evolutionary sequence: Type 1 $\to$ intermediate type $\to$ Type 2 as the accretion rate decrease. On the other hand, there are still some distances between the off-state points and the critical line. Correspondingly, some spectra from the ``turn-off" state still contain a broad $H\alpha$ component and, most likely, a featureless continuum. The analysis results from STARLIGHT also suggest that there are some contributions from AGN emissions for most ``turn-off" spectra. This result may suggest that although the central engine and BLR change substantially at ``turn-off" state, the BLR does not disappear completely in some ``turn-off" CL-AGNs.

Our results show again that the changes in dust extinction cannot account for the changes in the broad emission lines, and support that the CL-AGNs may be associated with the instability of accretion disk. Following the decrease in the accretion rate, the BLR disappears (though not completely in some cases), and the featureless continuum becomes much weaker. 

\subsection{Stellar Population vs. BH parameters}

CL-AGNs provide a valuable opportunity to directly obtain both BH properties (on-state, when the AGN dominates)  and host galaxies properties (off-state, when the AGN fades). 

The stellar populations of CL-AGNs and the comparisons with normal AGNs (Type 1s and Type 2s) have been summarized in  Section ~\ref{sec:stellar_population}. There is no significant difference in stellar populations between CL-AGNs and normal AGNs. All of them have past star-formation activities. However, the host galaxies of CL-AGNs have relatively more contributions from intermediate populations, while Type 1 AGN hosts have more contributions from young stellar populations, and in Type 2 hosts the old stellar populations contribute slightly more. If this analysis is correct, it means that the stellar populations of host galaxies may be related to the spectral types of their central AGNs, which is consistent with the coevolutionary sequence of AGNs and their hosts \citep{2007MNRAS.382.1415S,2008ApJ...673..715C,2009ApJ...705.1336C}. 

However, there are some uncertainties in this result. Firstly, for Type 1 AGNs, the contributions from young stellar populations may be overestimated due to the SB-FC degeneracy. Secondly, Type 2 AGNs may be composed of two populations: one is due to the well-known orientation effect (intrinsically the same as Type 1 AGNs), and the other is the intrinsically low accretion rate due to the absence of BLR (``true" Type 2 AGNs). However, there has been no study on the stellar population difference between these two kinds of Type 2 AGNs.  It is necessary to analyze the stellar populations of these two kinds of Type 2 AGNs separately, which will help us to better understand the real position of CL-AGNs in the BH parameter space.

\begin{figure}[!htb]
	\begin{center}	
		\includegraphics[page=1,scale=0.37]{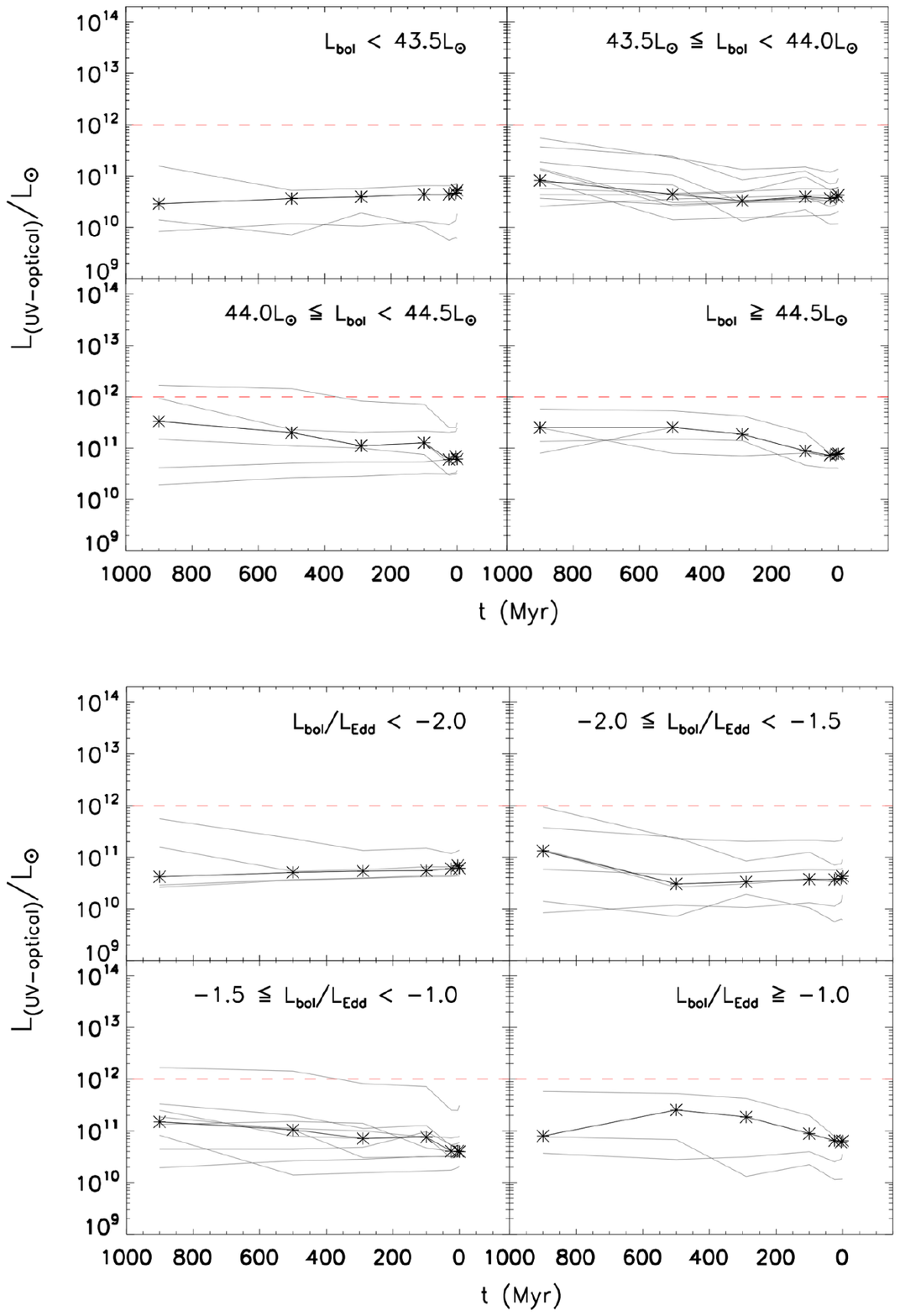}   
		\caption{The average $\rm L_{UV-optical} (912-9000$ $\rm \AA)$ for $\rm L_{bol}$ bins (upper panel) and $\rm L_{bol}/L_{Edd}$ bins (lower panel) of CL-AGNs. Past star-formation activity tends to be more active in source which shows more AGN activity (higher  $\rm L_{bol}$ or  $\rm L_{bol}/L_{Edd}$).}
	\end{center}
	\label{fig:SFH_com} 
\end{figure}

In order to further investigate the correlation between host galaxies and BH properties, we compare the SFH under different AGN activity levels in our CL-AGN sample. We bin the $\rm L_{bol}$ (or $\rm L_{bol}/L_{Edd}$) into four bins ($\rm L_{bol} < 43.5{L_{\sun}}$, $\rm 43.5{L_{\sun}}\le L_{bol} < 44.0{L_{\sun}}$, $\rm 44.0{L_{\sun}} \le L_{bol} < 44.5{L_{\sun}}$, $\rm L_{bol} \ge 44.5{L_{\sun}}$). Both $\rm L_{bol}$ and $\rm L_{bol}/L_{Edd}$ are calculated from ``turn-on" state, and the SFHs are obtained from the ``turn-off" state. Therefore, the correlation is unlikely due to the SB-FC degeneracy, where the young stellar populations are more likely to be overestimated when there are more emissions from central AGN. Through the comparisons in Figure ~\ref{fig:SFH_com}, we find that the CL-AGNs with higher $\rm L_{bol}$ (or $\rm L_{bol}/L_{Edd}$) tend to have more active star-formation activities in the past. This implies that the past star-formation activities are indeed associated with the AGN physical properties, and, as a result, the large-scale environments of CL-AGN hosts may be an important factor related to the CL phenomenon. As mentioned before, some recent studies on CL-AGN hosts have demonstrated a potential link between the CL-AGN host galaxies and their CL phenomena \citep{2019ApJ...876...75C,2019MNRAS.486..123R,2020MNRAS.498.3985Y}. 
Post-SB activities were also found in CL-AGNs. In \cite{2013ApJ...762...90C}, the CL-AGN J210200.42+000501.8 was classified as a post-SB quasar, and CL-AGN J101152.98+544206.4 was found to be a post-SB galaxy \citep{2016ApJ...821...33R}.  Combining our results with previous works, we suggest that there may be a potential link between the host environments of CL-AGNs and their changing-look phenomena, and further studies on this link are needed.

\section{Summary}

In this work, we investigate the stellar populations of 26 CL-AGNs, which are collected from the literature and all have SDSS ``turn-off" spectra. The main results are as follows:

(1) Using the stellar population synthesis code STARLIGHT, we determine the detailed stellar populations of CL-AGN host galaxies for the first time, and derive the SFHs for them. Generally, the stellar populations of CL-AGNs lie between SB and normal galaxies, which is consistent with previous results showing that there is no significant difference between CL-AGNs and NCL-AGNs. However, when a detailed comparison is made between the AGN hosts themselves, the CL-AGN hosts have relatively more contributions from the intermediate stellar populations.

(2) We estimate the stellar velocity dispersion ($\rm \sigma_{\star}$) by fitting ``turn-off" spectra, and derive the virial BH mass ($\rm M_{BH,vir}$) using the $\rm R_{BLR}$-$\rm L_{5100}$ relation, with the values obtained from ``turn-on" spectra. Combined with previous studies, we suggest that the CL-AGNs follow the overall $\rm M_{BH}$-$\rm \sigma_{\star}$ relation.

(3) The decomposition of the multi-epoch spectra and the dereddened fitting results of the continuum and broad emission lines in the $\rm H\alpha$ region disfavor the dust extinction scenario for CL-AGNs.

(4) We compare the distributions of CL-AGNs in the BH parameter space ($\rm L_{bol}$ versus $\rm L_{M_{BH}}$, $\rm L_{bol}$ versus $\rm L_{bol}/L_{Edd}$) for both ``turn-on" and ``turn-off" states. This comparison shows that CL-AGNs in ``turn-on" states have lower Eddington ratios relative to the overall quasars, and that ``turn-off" state are located near the critical region below which the accretion rate is barely enough to support a BLR. The BPT diagrams also indicate that CL-AGNs lack strong accretion signatures. All these results support the variable accretion rate and the disk-wind scenario.

(5) There may be a potential link between the SFHs of CL-AGN hosts and their central BH properties. Firstly, the CL-AGNs whose central BHs are more active (with higher $\rm L_{bol}$ or $\rm L_{bol}/L_{Edd}$) at present tend to have more active star-formation activities in the past. Secondly, the stellar populations of CL-AGNs may be related to the AGN spectral type.  However, the contamination from the central engine in Type 1 AGNs, and the lack of distinction between the two kinds of Type 2 AGNs (one based on the unification model, and the another due to the intrinsically low accretion rate), may bring uncertainties to this result. A further analysis of the stellar populations of the two kinds of Type 2 AGNs is necessary.

\acknowledgements

We thank the anonymous referee for providing very insightful comments and helpful suggestions. The authors are grateful for the support of the National Key R\&D Program of China (2016YFA0400703), the National
Science Foundation of China (grants 11533001, 11721303, and 11927804) and the China Postdoctoral Science Foundation (2021M690229).

This publication makes use of the SDSS database and the DR7 edition of the SDSS quasars catalog. Funding for the SDSS has been provided by the Alfred P. Sloan Foundation, the Participating Institutions, the National Science Foundation, the U.S. Department of Energy,
the National Aeronautics and Space Administration, the Japanese Monbukagakusho, the
Max Planck Society, and the Higher Education Funding Council for England. 

The Guoshoujing Telescope (the Large Sky Area Multi-object Fiber Spectro- scopic Telescope LAMOST) is a National Major Scientific Project built by the Chinese Academy of Sciences. Funding for the project has been provided by the National Development and Reform Commission. LAMOST is operated and managed by the National Astronomical Observatories, Chinese Academy of Sciences. 

\clearpage

\appendix
\renewcommand{\appendixname}{Appendix ~\Alph{section}}
\section{``turn-on" and ``turn-off" spectra}
\label{Appendix:Specs}

In Figure ~\ref{fig:Specs}1, we present both the ``turn-on" and the ``turn-off" spectra for all of the CL-AGNs in our sample, except for J0158-0052 which only has a ``turn-off" spectrum.


\includegraphics[page=1,scale=0.65]{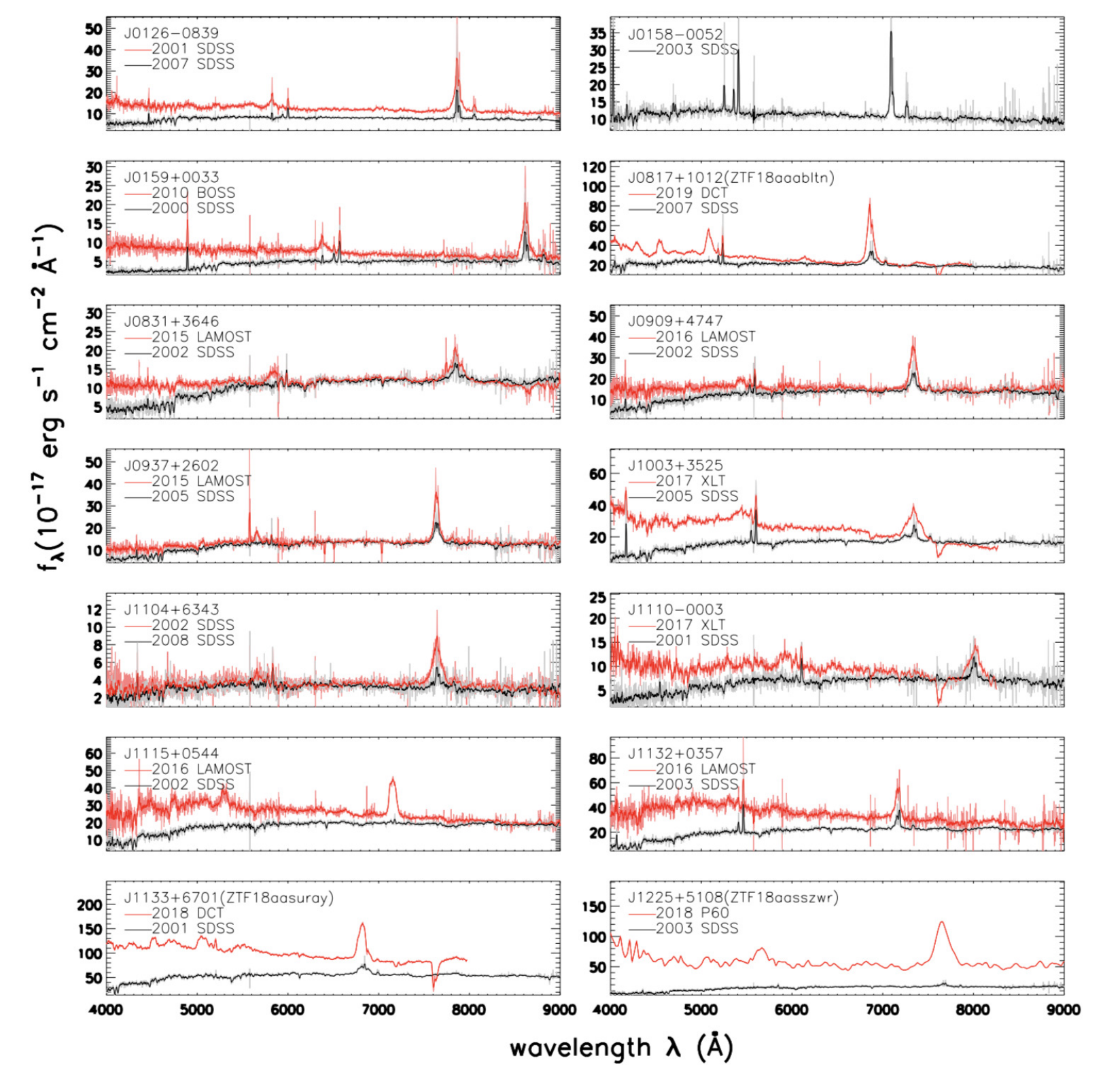} 
\normalsize{Figure A1. ``turn-on" (red) and ``turn-off" (black) spectra (4000-800 $\rm \AA$) for 26 CL-AGNs in our sample. The sources are listed in an order of their coordinates.  No corrections for the redshift and Galactic extinction. The object name, the instruments used to observe and the date of observations are denoted at the top-left corner.}
\label{fig:Specs}
\clearpage 

\includegraphics[page=1,scale=0.58]{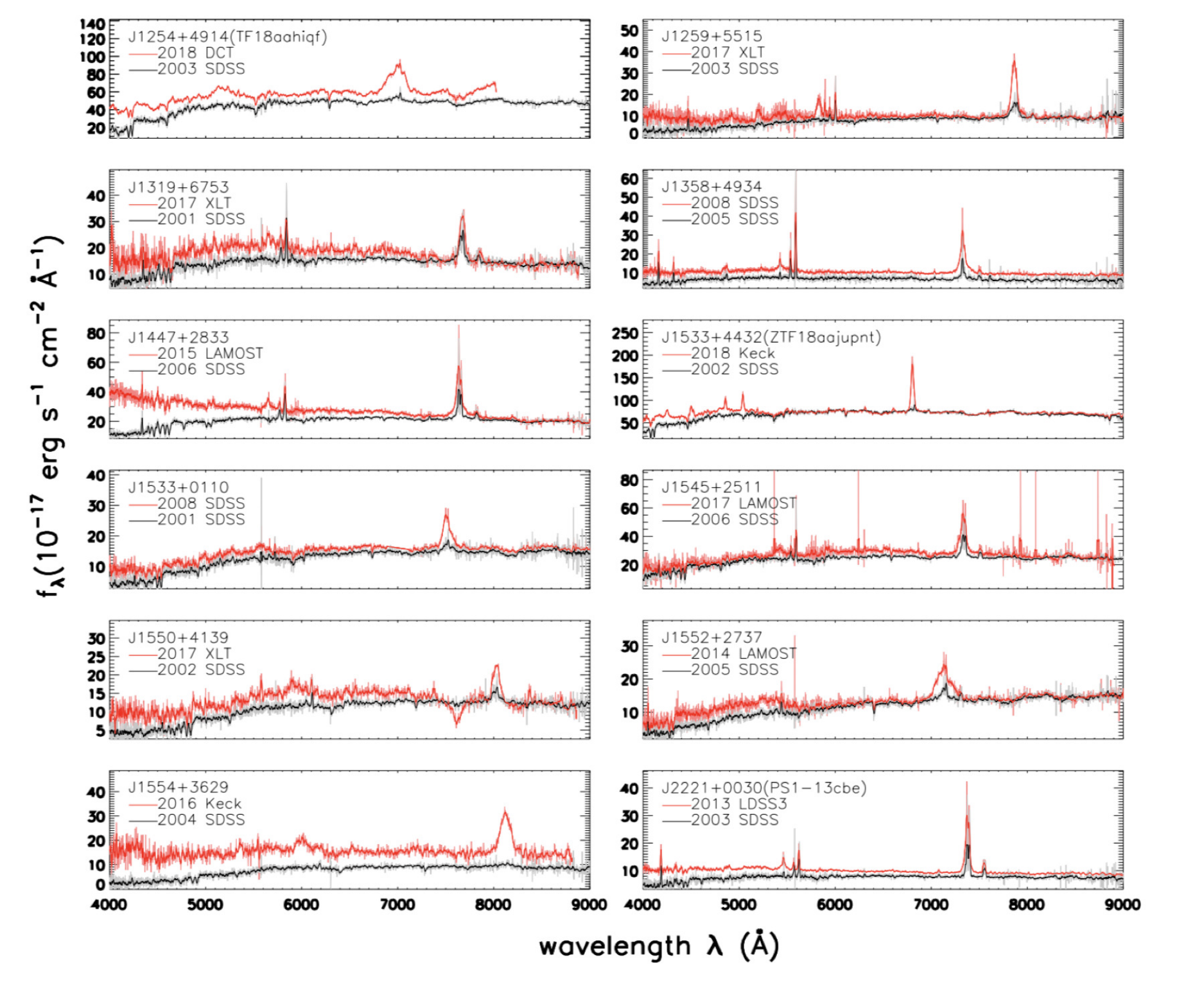} 
\center{\normalsize{Figure A1. -continued}}

\end{document}